\documentclass[aps,prx,twocolumn,notitlepage,nofootinbib]{revtex4-1}
\usepackage{epsfig,amssymb,amsmath,mathrsfs,amsthm,graphicx,float,xcolor,hyperref}

\usepackage{comment}
\usepackage{subfigure}
\usepackage{algorithm}
\usepackage{algpseudocode}
\usepackage{refcount}

\newcommand{\N}{\mathbb{N}}

\newcommand{\set}[1]{\mathsf{#1}}

\newcommand{\spc}[1]{\mathcal{#1}}

\def\d{{\rm d}}

\newcommand{\Lin}{\mathsf{Lin}}

\def\>{\rangle}
\def\<{\langle}

\newcommand{\map}[1]{\mathcal{#1}}
\newcommand{\Tr}{\operatorname{Tr}}

\floatname{algorithm}{Protocol}

\newtheorem{theo}{Theorem}

\newtheorem{lemma}{Lemma}

\newtheorem{cor}{Corollary}
\newtheorem{defi}{Definition}
\newtheorem{rem}{Remark}

\def\qed{$\blacksquare$ \newline}

\begin{document}

\title{Accuracy enhancing protocols for quantum clocks} 

\author{Yuxiang Yang} \affiliation{Institute for Theoretical Physics, ETH Z\"urich, Switzerland}
\author{Lennart Baumg\"artner} \affiliation{Institute for Theoretical Physics, ETH Z\"urich, Switzerland}
\author{Ralph Silva} \affiliation{Institute for Theoretical Physics, ETH Z\"urich, Switzerland}
\author{Renato Renner} \affiliation{Institute for Theoretical Physics, ETH Z\"urich, Switzerland}

\begin{abstract}
The accuracy of the time information generated by clocks can be enhanced by allowing them to communicate with each other. Here we consider a basic scenario where a quantum clock receives a low-accuracy time signal as input and ask whether it can generate an output of higher accuracy. We propose protocols that use a quantum  clock with a $d$-dimensional state space to achieve an accuracy enhancement by a factor of $d$, for large enough $d$. If no feedback to the input signal is allowed, this enhancement is temporary. With feedback the accuracy enhancement can be retained indefinitely. Our protocols are specific to quantum clocks, and may be used to synchronise them in a network, defining a time scale that is more accurate than what can be achieved by non-interacting or classical clocks.
\end{abstract}

\maketitle

\section{Introduction}
That progress in quantum technologies is commonly accompanied by progress in high-precision time-keeping, as witnessed again recently~\cite{bloom2014optical,ludlow2015optical,ludlow2018optical}, is not a coincidence. There are indeed fundamental reasons  why the use of quantum phenomena enables more accurate time measurements than purely classical means~\cite{Bollinger1996,Huelga1997}. One of these reasons is of information-theoretic nature | a quantum clock with a $d$-dimensional state space can hold $\log_2 d$ qubits of information about time, whereas a corresponding classical clock only holds $\log_2 d$ classical bits.   As shown in~\cite{woods2018quantum}, this makes a difference. 
A quantum clock can achieve an accuracy (almost) quadratic in~$d$, whereas the accuracy of any classical clock is always bounded by~$d$.\footnote{This statement refers to the accuracy measure $R$ discussed in Section~\ref{sec_accuracymeasure} below.} 

From the viewpoint of information theory, it is natural to not only study clocks individually, but rather consider scenarios where multiple clocks  can communicate with each other and hence exchange information about time. This is practically relevant, since networks of clocks are commonly used to define a time scale that is more reliable than what any individual clock could achieve.\footnote{The International Atomic Time, which serves as a basis for the Coordinated Universal Time (UTC),  is defined as an average of the reading of approximately 400 atomic clocks, with a weighting that depends on the measured stability of the individual clocks.} Furthermore, it is known that the accuracy of frequency measurements can be enhanced with correlated quantum systems~\cite{Bollinger1996,Huelga1997}.  In future networks of quantum clocks, this fact may be exploited to define a highly accurate global time scale~\cite{komar2014quantum}. 

\begin{figure}[b!]
\begin{center}
  \includegraphics[width=\linewidth]{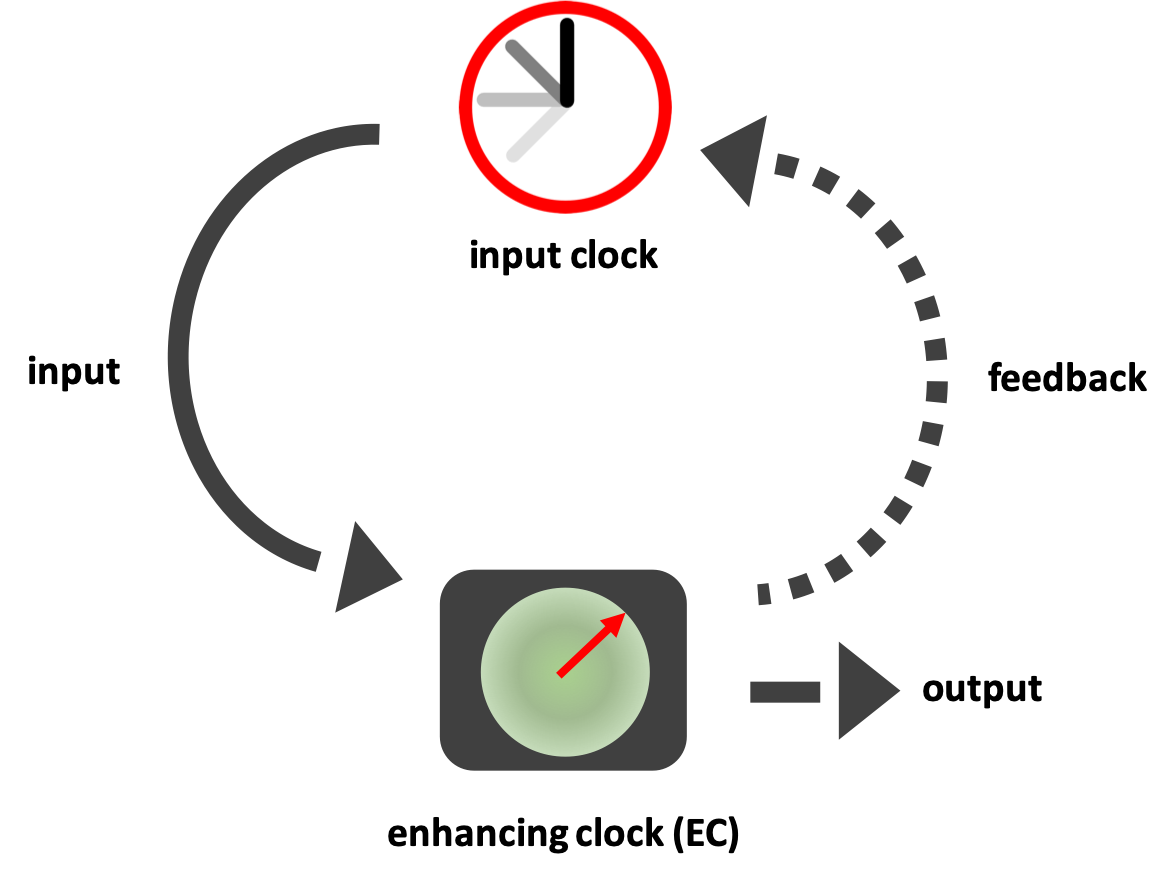}
  \end{center}
  \caption{{\bf Accuracy enhancement.} The enhancing clock (EC) receives a clock signal from an input clock and produces an enhanced clock signal as output. The performance of the protocol can be further improved when feedback to the input clock is allowed. }\label{fig:scheme}
\end{figure}

Here we study a basic task that a clock may carry out within such a communication scenario: the enhancement of time information (see Fig.~\ref{fig:scheme}). More precisely, we consider a setup where one clock, the \emph{Enhancing Clock (EC)}, receives information from another clock about what time it is. Combining this input with internal information, the EC is supposed to output more accurate information about what time it is. For this, it may also send feedback to the clock that generates the input.  We note that, while this scenario merely involves two clocks, it serves as a building block for larger clock networks. 
 
The concept of enhancing time information also plays a crucial role for operating individual high-precision clocks, such as atomic clocks~\cite{derevianko2011colloquium,ludlow2015optical,ludlow2018optical}.  In a caesium clock, for instance, one may regard the gas of caesium atoms as the EC, which receives an input, in the form of a microwave signal, from an electronic oscillator (e.g., a crystal oscillator), and also feedbacks to this oscillator.  Radio clocks also fall into this scheme. They receive time signals sporadically (e.g., once an hour), which they use in combination with an internal quartz crystal to output a continuous time signal. 

As already noted, the ability of  clocks to generate accurate time information is related to their size, measured in terms of the dimension~$d$ of their state space | the larger $d$ is, the more accurate the clock can be~\cite{rankovic2015quantum,woods2018quantum}.  Practically, the parameter $d$ describes the size of the system that one has control over and, eventually, determines the cost of constructing the clocks. Fundamentally, the dimension captures the information capacity of any system and is regarded as one of the most fundamental resources of physics; see, for instance, the Bekenstein bound \cite{bekenstein1981universal}.  Here we show that $d$ is also the relevant parameter for the task of enhancing time information. Specifically, we propose protocols which allow an EC of size $d$ to enhance the accuracy of an input signal up to a factor of $d$. These protocols make use of quantum effects, allowing them to outperform basic classical protocols.

The remainder of this paper is structured as follows.  Section~\ref{sec_accuracymeasure} is devoted to the modelling of clocks and the question of how to operationally quantify their accuracy.  In Section~\ref{sec-nofeedback} we describe and analyse a basic protocol for accuracy enhancement by quantum clocks. We extend this protocol in Section~\ref{sec-feedback} to include feedback. In Section~\ref{sec-incoherent}, we explain why the EC in our protocols has to be a quantum clock and compare our protocols to those that process input ticks without making use of quantum dynamics. 
In Section~\ref{sec-network}, we show a possible application of our protocols to establishing a shared clock signal in a network.
Finally, in Section~\ref{sec-discussion} we  conclude with a discussion on other possible applications of our approach.

\section{Quantifying Accuracy}  \label{sec_accuracymeasure}

For our purposes, a clock is a quantum system that autonomously emits information about time.\footnote{One may distinguish between two types of time keeping devices. ``Stopwatches''  measure an interval of time determined by external events (e.g., when we press a button). Conversely, ``reference clocks'' autonomously generate a reference frame for time, i.e., they themselves trigger events  (e.g., the ringing of a bell tower). Here we are concerned with the latter.} We suppose that this time information comes in the form of \emph{ticks}, which subdivide time into intervals~\cite{rankovic2015quantum}. For a perfectly accurate clock, the emitted ticks are perfectly regular, so that the intervals defined by them all have equal length. The accuracy of a clock can thus be quantified in terms of how much it  deviates from this ideal.  

Before proceeding with the definition of an accuracy measure, we point out the conceptual distinctions between this approach and the treatment of clocks in the context of metrology, which is common in the literature (see, for instance, Refs.\ \cite{buvzek1999optimal,jozsa2000quantum,komar2014quantum}). In quantum metrology, a clock is modelled as a  system that oscillates at a frequency~$\omega$. One then considers the task of estimating~$\omega$, or more generally the difference between $\omega$ and the frequency of an external oscillator, by interrogating this system, i.e., one prepares it in a given state, lets it evolve, and then measures it. This  measurement is optimised to minimise the uncertainty in the estimate for~$\omega$~\cite{Bollinger1996,Huelga1997}. In contrast, within this work we follow~\cite{rankovic2015quantum,erker2017autonomous,woods2018autonomous,woods2018quantum} and regard a clock as an autonomous device that directly outputs information about what time it is, rather than a frequency.\footnote{Converting an estimate for the frequency~$\omega$ of an oscillator into information about what time it is requires additional resources.  Bounds on the accuracy of the former  do hence not directly translate into bounds on the accuracy of the latter, nor vice versa.}  Furthermore, time information is continuously emitted as the clock is running, rather than at the end of an  external process.

To motivate the definition that follows for an accuracy measure operationally, we note that the ticks emitted by a clock can be used to time-tag events. The tick that is generated before all the others is taken as a starting point, and we therefore refer to it as the $0$-th tick. If an event occurred before the $j$-th tick, we tag it with ``$j-1$'', and if it occurs after, we tag it with ``$j$''.  For a perfect clock, whose ticks occur at fixed times, this tagging is deterministic and, in this sense, unambiguous. This is no longer the case for an imperfect clock, where the exact emission time $T_j$ of the $j$-th tick can be random. Nevertheless, we may define a $(1-\epsilon)$-confidence interval  $\set{C}_{j}$ of time for each tick~$j$, demanding that $\set{C}_{j}$ contains $T_j$ except with a fixed probability $\epsilon > 0$. Whenever an event lies either before or after the interval~$\set{C}_{j}$, the distinction between tag ``$j-1$'' and ``$j$'' would still be unambiguous with probability at least~$1-\epsilon$. Only events that occur within the interval $\set{C}_{j}$ have a higher chance of being classified erroneously.

Following this idea, we may introduce a family of inaccuracy measures, $\Sigma_{j}(\epsilon)$, parameterised by $\epsilon \in [0,1]$ and $j \in \mathbb{N}$.  A first attempt is to take $\Sigma_{j}(\epsilon)$ to be the width of $\set{C}_{j}$. However, this choice of the inaccuracy measure, as well as other common measures like the standard deviation, are not dimensionless. Since the clocks we considered are autonomous, their inaccuracy should still be quantifiable without external reference. Therefore, the measure of inaccuracy should be \emph{absolute} and free of the unit of time. In this spirit, we construct a unit-free measure by dividing the width of $\set{C}_{j}$ by the distance between confidence intervals. Roughly, $\Sigma_{j}$ can be interpreted as the fraction of  events for which the time tag~``$j$'' may be ambiguous, i.e., events that lie within the interval $\set{C}_{j}$, among all events that occur between the $(j-1)$-th and the $j$-th tick.

\begin{defi}[Inaccuracy]
For any desired \emph{confidence level} $1-\epsilon$, the \emph{$\epsilon$-inaccuracy of the $j$-th tick} is  defined as
\begin{align}\label{accuracy}
\Sigma_{j}(\epsilon):=\inf_{\substack{\set{C}_j = [\mu-\frac{\sigma}{2}, \mu+\frac{\sigma}{2}] \\ \Pr[T_j \notin \set{C}_j] \leq \epsilon }} \frac{\sigma}{\mu/j} \ ,
\end{align} 
where the infimum ranges over intervals $\set{C}_j$ with any width~$\sigma$ and centre~$\mu$ that contain the time $T_j$ of the $j$-th tick with probability at least $1-\epsilon$. 
\end{defi} 


An important special case is that of an \emph{i.i.d.}\ clock, where the time durations between ticks,  i.e., the differences $T_{j}-T_{j-1}$, are \emph{independent and identically distributed} for all $j \in \mathbb{N}$. Setting $T_0 = 0$, we then have  $P_{T_{j}-T_{j-1}} = P_{T_1}$ for any $j \in \mathbb{N}$. The behaviour of an i.i.d.\ clock is thus fully defined by the distribution of the first tick $T_1$. We will therefore in the following often write $T$ instead of $T_1$, and analogously $\Sigma(\epsilon)$ instead of $\Sigma_1(\epsilon)$. For i.i.d.\ clocks, the distance between the confidence intervals of $T_j$ and $T_{j-1}$ is of the order $\mu/j$. The inaccuracy $\Sigma(\epsilon)$ thus has the operational interpretation that when it becomes larger than $1$, the confidence intervals of successive ticks begin to overlap, i.e., the confidence level of time-tagging has dropped below $1-\epsilon$ for all events.

For an i.i.d.\ clock one can use the Hoeffding inequality to obtain a bound on the $\epsilon$-inaccuracy of its ticks.  Suppose that the first tick lies in the interval $\set{C}_1 = [\mu_1-\frac{\sigma_1}{2}, \mu_1 + \frac{\sigma_1}{2}]$ with probability at least $1-\epsilon$. Suppose furthermore that $\Sigma(\epsilon) = \frac{\sigma_1}{\mu_1} \leq 1$. Then, for any $n \in \mathbb{R}^+$, Hoeffding's inequality implies that the probability that the $j$-th tick misses an appropriately centred interval $\set{C}_j$ of size $\sigma_{j}= n \sqrt{j} \sigma_1$ is at most 
\begin{align} \label{iid-epsilon}
  \epsilon_{j,n}:=1-(1-\epsilon)^j(1-2e^{-n^2/2}).
\end{align}
We thus have
\begin{align}\label{iid-accuracy}
\Sigma_{j}(\epsilon_{j,n})\le 2n\cdot\sqrt{j}\cdot  \Sigma(\epsilon).
\end{align}
Note that the confidence $1-\epsilon_{j,n}$ decreases monotonically. To obtain a bound that holds for a constant confidence $1-\epsilon$, we need to know more about the tail of the distribution of $T$. For example, if it is sub-Gaussian (meaning that it has a tail that vanishes at least as fast as some Gaussian distribution),
 one can employ Hoeffding's inequality for sub-Gaussian random variables to obtain the inaccuracy bound \cite[Theorem 2.6.2]{vershynin2018high}
\begin{align}\label{subgaussian}
\Sigma_{j}(\epsilon)\le c\cdot \sqrt{j\cdot\ln\left(\frac{2}\epsilon\right)},
\end{align}
where $c>0$ is a constant that depends only on the distribution of $T$, but not  on $j$ and $\epsilon$. Many types of tick distributions including Gaussian distributions themselves as well as any distributions with finite support are sub-Gaussian. Another example is a situation where the distribution is a mixture of several perfect tick signals (i.e., delta functions) with slightly different frequencies.
 Eq.~(\ref{subgaussian}) holds for all these cases, and $\Sigma_j(\epsilon)$ grows like $\sqrt{j}$ for any fixed $\epsilon$. 

Our inaccuracy measure can be compared to another measure of clock accuracy considered in \cite{erker2017autonomous,woods2018quantum}, which is defined as $R_j:=\mathrm{E}(T_j)^2 / \mathrm{Var}(T_j)$, with $\mathrm{E}(T_j)$ and $\mathrm{Var}(T_j)$ being the mean and the standard deviation of $T_j$, respectively. 
The latter is sensitive to the tail behavior of $T_j$'s probability distribution, i.e., ticks that occur with small probability but deviate a lot from the mean. The value $\Sigma_j(\epsilon)$ for fixed $\epsilon > 0$ does hence not imply a bound on $R_j$. Conversely, for any clock tick achieving accuracy $R_j$ in terms of the measure considered in \cite{erker2017autonomous,woods2018quantum}, it is immediate from Chebyshev's inequality that it has an $\epsilon$-inaccuracy
\begin{align}
\Sigma_j(\epsilon)\le\sqrt{\frac{j^2}{\epsilon\cdot R_j}}.
\end{align}
Furthermore, for i.i.d.\ clocks the accuracy measure satisfies $R_j=j\cdot R_1$, so that we have
\begin{align}
\Sigma_{j}(\epsilon)\le \sqrt{\frac{j}{\epsilon\cdot R_1}}.
\end{align}

\section{Basic accuracy enhancing}\label{sec-nofeedback} 

Given an input clock signal, the goal of accuracy enhancement is to produce an output signal with as small inaccuracy as possible, using an enhancing clock (EC) as in Fig.\ \ref{fig:scheme}. To model the EC, we use the concept of autonomous clocks developed in~\cite{erker2014quantum,rankovic2015quantum, erker2017autonomous,woods2018autonomous}, which produce signals without an external time reference. 
 This model of clocks originates naturally from open system dynamics \cite{breuer2002theory}, in particular the theory of quantum trajectories \cite{brun2002simple}. The evolution of the clock is a time-homogeneous Markov process that is effectively the joint action of a unitary evolution and a time-independent weak measurement repeated at very high frequency.  
An autonomous clock is thus characterized by two key ingredients: a finite-dimensional clock system (which can be either classical or quantum) that evolves continuously in time and a \textit{detector} that \emph{constantly} measures the clock system and produces ticks \cite{rankovic2015quantum,woods2018quantum}.  In what follows, we will usually operate the clock as a \emph{reset} clock as in~\cite{woods2018quantum}, i.e., it returns to the same state, called the \emph{reset state}, after each tick.

We take the EC to be  an autonomous quantum  clock together with a switch that determines whether the detector is off or on. The clock is designed so that if the detector is off, the dynamics of the clock, denoted by $\map{D}_{\rm no-tick}$, is unitary and periodic. If the detector is on, the dynamics of the clock, denoted by $\map{D}_{\rm tick}$, corresponds to that of the autonomous clock as defined in~\cite{woods2018quantum}, i.e.,  the clock state is constantly measured and ticks can be produced.

We now introduce a \emph{stability criterion} for such an EC, or, more precisely, a family of ECs parameterised by their dimension~$d$, indicated as ${\rm EC}_d$ hereafter. The criterion will play a crucial role for the formulation of our main results. Suppose that the EC first evolves according to $\map{D}_{\rm no-tick}$ and is then, at some time, switched to $\map{D}_{\rm tick}$. Since the dynamics $\map{D}_{\rm no-tick}$ is periodic, we can label the state by a time parameter $s\in(-\tau^{{\rm EC}_d}/2,\tau^{{\rm EC}_d}/2]$, where $\tau^{{\rm EC}_d}$ is the period of the EC when it evolves according to $\map{D}_{\rm no-tick}$. We choose the reference for $s$ such that $s=0$ corresponds to the clock being in the reset state. For any switching time $s$, let $\smash{T^{{\rm EC}_d}_s}$ be the additional time it takes until the EC emits a tick. Therefore the time parameter corresponding to the tick is $\smash{T^{{\rm EC}_d}_s} + s$. A stable clock is one for which this parameter is independent of the choice of $s$. The precise definition is as follows.


 \begin{figure}[b!]
\begin{center}
  \includegraphics[width=\linewidth]{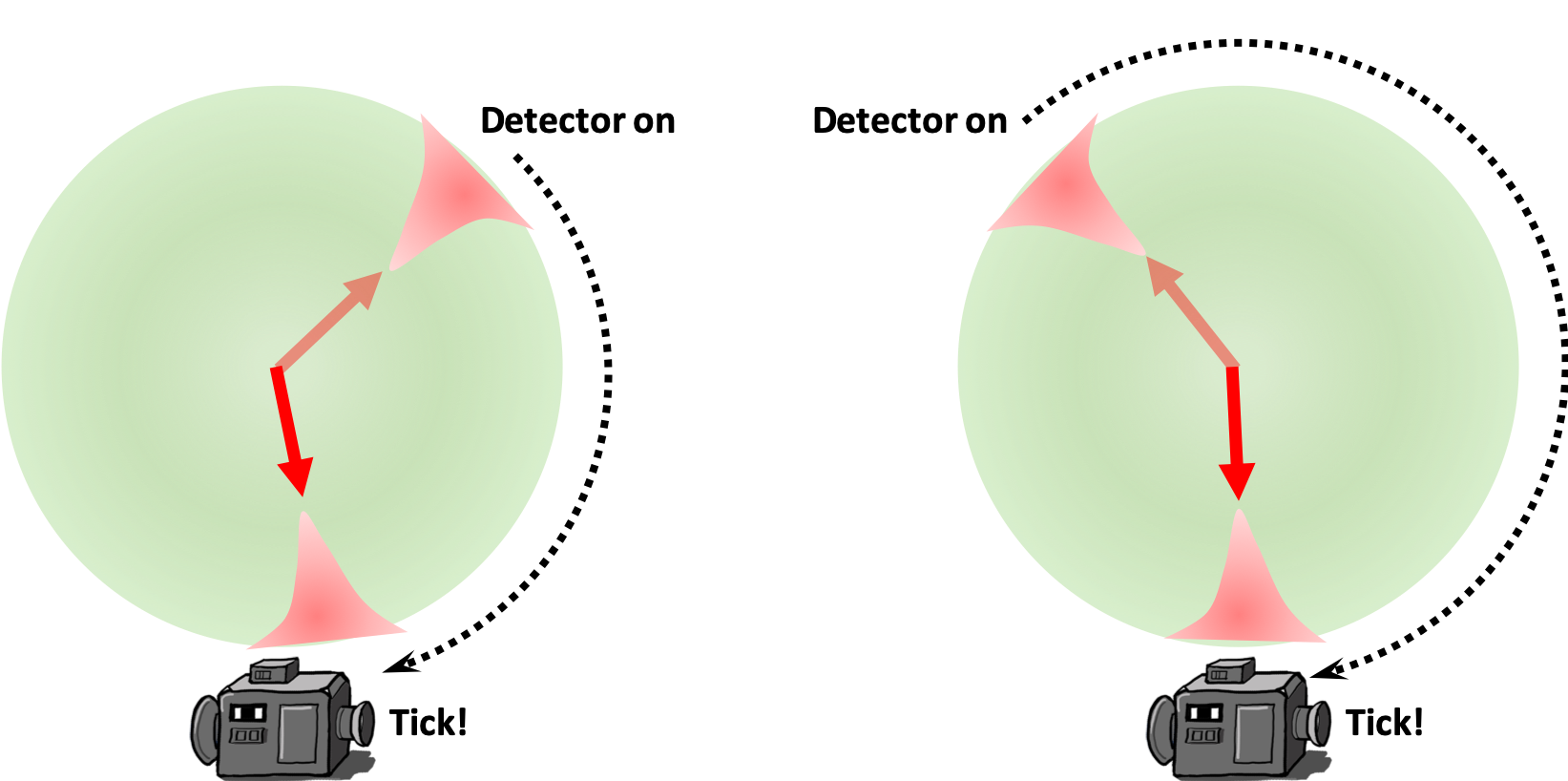}
  \end{center}
  \caption{{\bf A stable enhancing clock.} An EC satisfying the stability criterion (see Definition \ref{defi-clock}) is illustrated in the above figure. When the dimension $d$ of the EC grows large, the uncertainty of the EC's tick is always bounded by the same quantity no matter when the detector is switched on [cf.\ Eq.\ (\ref{tailbound-clock})], and it is almost sure that the tick is produced when the clock state comes close to the detector.
   }\label{fig:stable}
\end{figure}

\begin{defi}[Stability criterion]\label{defi-clock}
An EC satisfies the stability criterion if there exist sequences $\{\sigma^{{\rm EC}_d}\}_{d}$ and $\{\epsilon^{{\rm EC}_d}\}_d$  that vanish in the limit of $d\to\infty$, such that 
\begin{align}
\Pr\left[T^{{\rm EC}_d}_s+s\in\left({\textstyle \frac{\tau^{{\rm EC}_d}-\sigma^{{\rm EC}_d}}{2},\frac{\tau^{{\rm EC}_d}+\sigma^{{\rm EC}_d}}{2}}\right)\right]
\ge 1-\epsilon^{{\rm EC}_d}\label{tailbound-clock}
\end{align} 
holds for any $d$ and any $s\in \left(-\frac{\tau^{{\rm EC}_d}-\sigma^{{\rm EC}_d}}{2},\frac{\tau^{{\rm EC}_d}-\sigma^{{\rm EC}_d}}{2}\right)$.
\end{defi}
We stress that the stability criterion is defined only for ECs that evolve periodically under $\map{D}_{\rm no-tick}$, which means that the EC \emph{must be quantum} to satisfy this criterion (see Section \ref{sec-incoherent} for more details).  
As illustrated in Fig.~\ref{fig:stable}, one can imagine that a clock satisfying this criterion has a ``hand'' moving on the dial.  Eq.~(\ref{tailbound-clock}) then demands that the clock ticks if and only if the hand hits a detector located close to  $s=\tau^{{\rm EC}_d}/2$, regardless of (a)~how long the hand was evolving under $\map{D}_{\rm no-tick}$ and (b)~where the hand started, as long as it did not start too closely to the detector.

Note that the stability criterion is independent of the input signal and can thus be applied to any periodic autonomous quantum clock. If a clock satisfies the criterion then, when it is initialised in the reset state, its inaccuracy $\Sigma(\epsilon)$, for any $\epsilon > 0$, is upper bounded by
\begin{align}\label{R-c}
\bar{\Sigma}^{{\rm EC}_d}:=\frac{2\sigma^{{\rm EC}_d}}{\tau^{{\rm EC}_d}}
\end{align}
 for large enough $d$. 
 
The stability criterion is satisfied by \emph{Quasi-Ideal Clocks} \cite{woods2018autonomous,woods2018quantum} (see Lemma \ref{lemma-quasiideal} in the Appendix for details). These are the most accurate autonomous clocks for which analytical upper bounds on the inaccuracy have been calculated~\cite{woods2018quantum,woods2018autonomous}.  Specifically, a Quasi-Ideal Clock of dimension $d$ achieves a first-tick inaccuracy of $O(d^{-1+\nu})$ for any positive $\nu$ and for any confidence level $1-\epsilon < 1$  (see Lemma \ref{lemma-quasiideal} in Appendix~\ref{sec-proof-nofeedback}).\footnote{An analogous statement for the accuracy measure~$R_j$ has been proven in~\cite{woods2018quantum}.} 

 We now introduce our first protocol, which enhances the accuracy of input signals without using feedback to the input signal. The protocol requires a quantum EC satisfying the stability criterion in Definition \ref{defi-clock}.  
\begin{algorithm}[H]
  \caption{Accuracy enhancement without feedback by controlling the EC's switch.}
  \label{alg-nofeedback}
   \begin{algorithmic}[1]
   \Statex (Initialization) On receiving the first input tick, set the EC to the reset state and $\map{D}_{\rm tick}$.
   \Loop
   \State Wait for an EC tick.
   \State Produce an output tick, set the EC to the reset state and $\map{D}_{\rm no-tick}$.
   \State Wait for an input tick.
   \State Set the EC's dynamics to $\map{D}_{\rm tick}$.
   \EndLoop
   \end{algorithmic}
\end{algorithm}

\begin{figure}[t!]
\centering
\subfigure[]{\label{fig:goodcase}
 \includegraphics[width=\linewidth]{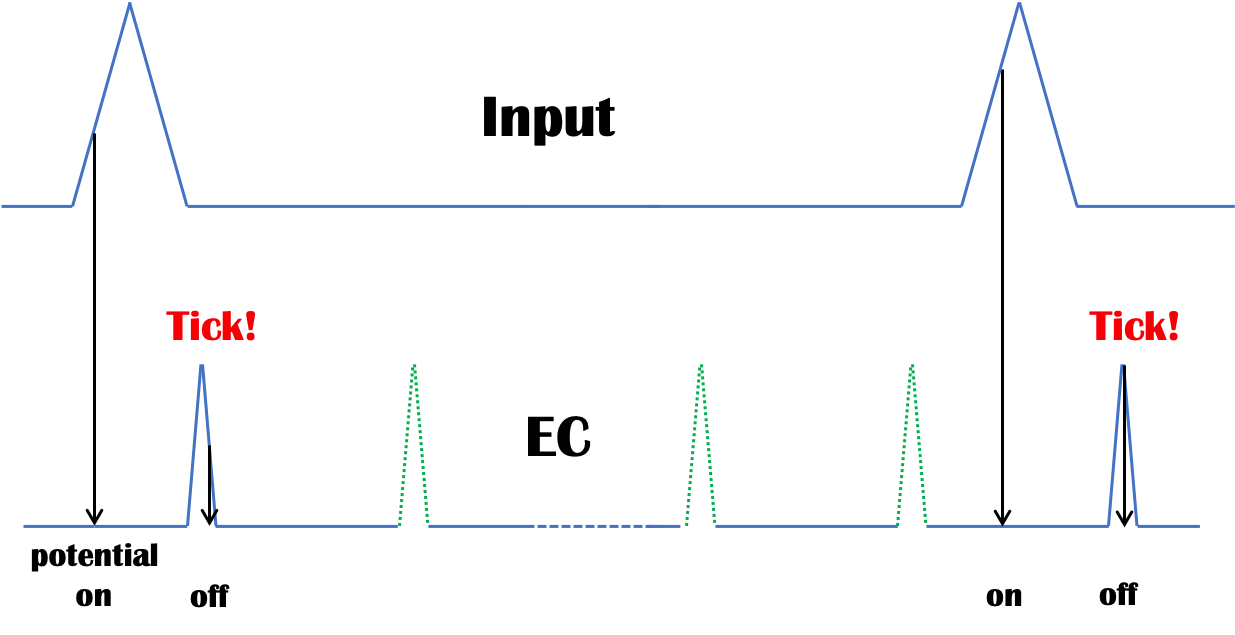}
}

\subfigure[]{\label{fig:badcase}
\includegraphics[width=\linewidth]{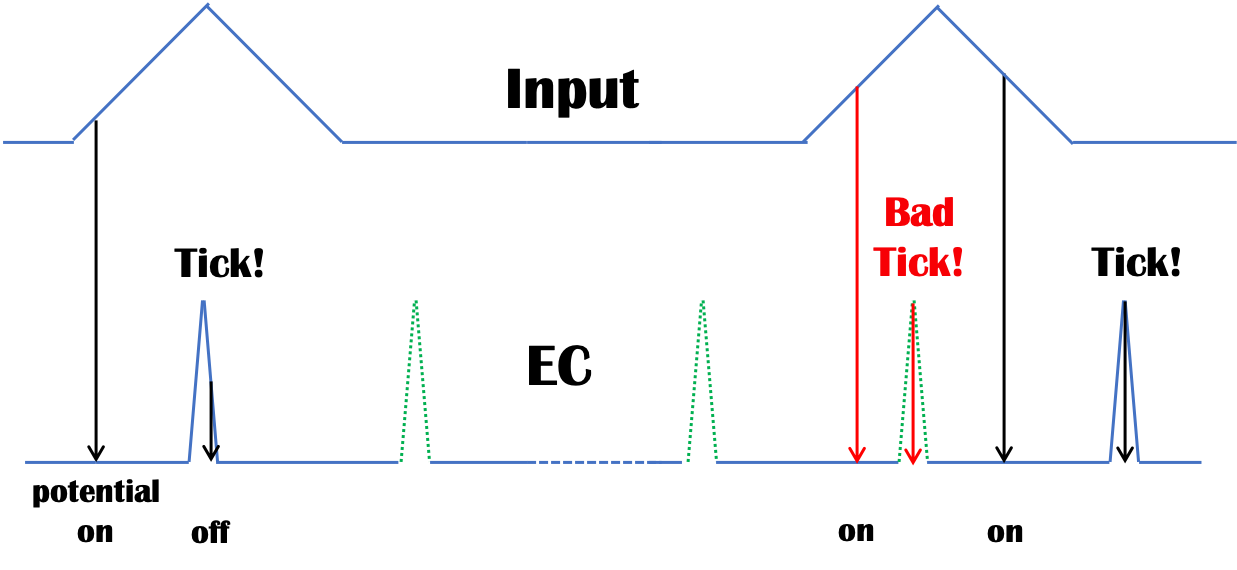}
}
\caption{{\bf Tick patterns of the input signal and the enhancing clock (EC) in Protocol~\ref{alg-nofeedback}.}  The spikes mark intervals within which ticks occur with high probability. Part~\subref{fig:goodcase} shows a situation where the protocol succeeds in generating accurate ticks, indicated in the lower half of the figure.  The ticks of the input signal, which are shown in the upper half of the figure, are used to turn on the detector of the EC. The green dotted spikes correspond to ticks of the EC that are suppressed when it evolves unitarily. Part~\subref{fig:badcase} shows the case where the width $\sigma^{{\rm in}}$ of the confidence intervals of the input ticks is larger than the period of the EC. Input ticks may then arrive too early, causing output ticks to be produced earlier than they should.}
\label{fig:ticks}
\end{figure}

To intuitively see why this protocol leads to accuracy enhancement, let us first consider an  i.i.d.\ input signal with a given $(1-\epsilon)$-confidence interval $\set{C}^{{\rm in}}$ of width $\sigma^{{\rm in}}$ and centered at $\mu^{{\rm in}}$. (Here and in the following, we will use superscripts ``${\rm in}$'', ``${\rm out}$'', or ``${\rm EC}_d$'' to refer to quantities characterising the input, the output, or the EC, respectively.) Generally speaking, the idea of the protocol is that  the output ticks should lie in a $(1-\epsilon_0)$-confidence interval $\set{C}^{{\rm out}}$ whose mean should be approximately equal to the mean of the input signal, i.e.,  $\mu^{{\rm out}}\approx\mu^{{\rm in}}$, whereas its width should scale like that of the enhancing clock, i.e. $\sigma^{{\rm out}}\approx\sigma^{{\rm EC}_d}$. Since, by Definition~\ref{defi-clock}, $\sigma^{{\rm EC}_d}$ is small for large $d$, this results in a reduction of the inaccuracy.

To achieve the accuracy enhancement, we choose the period of the EC to be slightly larger than $\sigma^{{\rm in}}$. As illustrated in Fig.\ \ref{fig:goodcase}, the EC will have its detector switched off after producing an output tick.
By its quantum nature, the EC evolves unitarily without any dissipation, until an input tick arrives and the detector is switched on. Next, by the stability criterion, the EC  always ticks when its clock state hits the detector, regardless of when the detector is switched on (see Definition \ref{defi-clock}). 
Fluctuations of the arrival time of the input tick, which result in an uncertainty of when the detector is switched on, will not affect the output ticks, and thus $\sigma^{{\rm out}}\approx\sigma^{{\rm EC}_d}$. It follows that the output inaccuracy can be expressed as $\Sigma^{{\rm out}}(\epsilon_0)\approx  \sigma^{{\rm EC}_d}/\mu^{{\rm in}} = \Sigma^{{\rm in}}(\epsilon)\cdot\left(\sigma^{{\rm EC}_d}/\sigma^{{\rm in}}\right)$, for any $\epsilon_0>\epsilon$ when $d$ is large enough. Here $\Sigma^{{\rm in}}=\sigma^{{\rm in}}/\mu^{{\rm in}}$ is the input inaccuracy.

However, the inaccuracy cannot be reduced arbitrarily by this method. Since the input tick is required to fall within the same period of the EC, the period $\tau^{{\rm EC}_d}$ should not be smaller than $\sigma^{{\rm in}}$. Otherwise, a bad tick with large deviation may be produced (see Fig. \ref{fig:badcase}). In addition, the probability for the output tick to lie in its confidence interval is at least the product of the probabilities that the input tick and the tick of the EC lie in their own confidence intervals, respectively, which means $\epsilon_0>\epsilon$. One finds that the minimal inaccuracy of our protocol's output signal is approximately the product of the inaccuracies of the input signal and the EC, i.e., 

\begin{align}
	\Sigma^{{\rm out}}(\epsilon_0)  \approx \Sigma^{{\rm in}}(\epsilon)\cdot \bar{\Sigma}^{{\rm EC}_d}
\end{align}
for $\epsilon_0>\epsilon$ and for large enough $d$. Here $\bar{\Sigma}^{{\rm EC}_d}$ is the inaccuracy upper bound of the EC given by Eq. (\ref{R-c}).


More precisely, we obtain the following lower bound on the accuracy of the output ticks; the proof can be found in Appendix \ref{sec-proof-nofeedback}.
 
\begin{theo}\label{theo-main}
Let the input clock be i.i.d.\ and such that $\Sigma^{{\rm in}}(\epsilon)<2/3$, and let the EC be such that the stability criterion holds. Then, for any $\epsilon_0>\epsilon$, for $j<2/(3\Sigma^{{\rm in}}(\epsilon))$, and for large enough $d$, the inaccuracy of the $j$-th output tick of Protocol~\ref{alg-nofeedback} satisfies
\begin{align}\label{out-bound-general}
\Sigma^{{\rm out}}_{j}(\epsilon_j)\le \frac{5j^2}{6}\cdot \Sigma^{{\rm in}}(\epsilon)\cdot \bar{\Sigma}^{{\rm EC}_d}\qquad \epsilon_j=j\cdot\epsilon_0.
\end{align}
\end{theo}

In particular, we can take the EC to be a Quasi-Ideal Clock, whose inaccuracy is nearly inversely proportional to $d$. This yields the following inaccuracy bound.

\begin{cor}\label{cor-nofeedback}
When the EC in Protocol \ref{alg-nofeedback} is taken to be a Quasi-Ideal Clock then, for any $\nu> 0$ and for large enough~$d$, 
\begin{align}\label{out-bound-quasiideal}
\Sigma^{{\rm out}}_{j}(\epsilon_j)\le \frac{5j^2}{3}\cdot \frac{\Sigma^{{\rm in}}(\epsilon)}{d^{1-\nu}}.
\end{align}
\end{cor}

The main result of this section, stated as Theorem \ref{theo-main} and Corollary \ref{cor-nofeedback}, is that Protocol \ref{alg-nofeedback} can reduce an input signal's inaccuracy almost by a factor equal to the dimension of the EC, given that the input signal is i.i.d. On the other hand, the right hand side terms of Eqs.~(\ref{out-bound-general}) and~(\ref{out-bound-quasiideal}) increase quadratically in~$j$, implying that the output signal becomes inaccurate faster than the i.i.d.\ signals [see Eq.~(\ref{iid-accuracy})]. The tail probability grows linearly in~$j$, like the input tick's tail probability before the accuracy enhancement [see Eq.~(\ref{iid-epsilon})]. Note that the period of the EC in the protocol must be chosen in accordance to the confidence interval of the input signal (see Appendix~\ref{sec-proof-nofeedback} for details).

\section{Accuracy enhancing with feedback}\label{sec-feedback}

As discussed above, the accuracy of the output ticks produced by Protocol~\ref{alg-nofeedback} drops after a certain time. It turns out that this problem can be remedied when feedback to the input clock is allowed.
The role of the feedback is to reset the input clock to its initial configuration in each round of producing a tick.

For any input signal generated by an input clock, the following protocol uses an EC that satisfies the stability criterion in Definition \ref{defi-clock} together with feedback to the input clock to produce more accurate output ticks.
\begin{algorithm}[H]
  \caption{Accuracy enhancement with feedback.}
  \label{alg-feedback}
   \begin{algorithmic}[1]
   \State (Initialization) On receiving the first input tick, set the EC to the reset state and $\map{D}_{\rm tick}$.
   \Loop
   \State Wait for an EC tick.
   \State Produce an output tick, set the EC to the reset state and $\map{D}_{\rm no-tick}$, and reset the input clock.
   \State Wait for an input tick.
   \State Set the EC's dynamics to $\map{D}_{\rm tick}$.
   \EndLoop
   \end{algorithmic}
\end{algorithm}

The working principle of this protocol is similar to that of the feedback-free protocol, and thus its performance is close to that of the feedback-free protocol in the short term.
The long-term stability follows from the fact that  the input clock is reset to its initial configuration after the production of an output tick. At the same time, the EC also resets.
Therefore, the output ticks are i.i.d., and the inaccuracy of the $j$-th output tick increases linearly in $\sqrt{j}$, as described by Eq.\ (\ref{iid-accuracy}).
To summarize, the performance of the feedback protocol is as good as the feedback-free protocol for the first few ticks and becomes higher as more output ticks are triggered, achieving long-term stability. A more detailed analysis leads to the following result. 

\begin{theo}\label{theo-feedback}
Let the input clock be such that $\Sigma^{{\rm in}}(\epsilon)<1$, and let the EC be such that the stability criterion holds. Then, for any $\epsilon_0>\epsilon$ and for large enough $d$, the output ticks of Protocol~\ref{alg-feedback} are i.i.d.\ with inaccuracy
\begin{align}\label{out-bound-general-feedback}
\Sigma^{{\rm out}}(\epsilon_0)< \Sigma^{{\rm in}}(\epsilon)\cdot\bar{\Sigma}^{{\rm EC}_d}.
\end{align}
\end{theo}

The proof of Theorem~\ref{theo-feedback} is provided in Appendix \ref{sec-proof-feedback}. As before, the EC can be taken to be a Quasi-Ideal Clock, yielding the following inaccuracy bound. 

\begin{cor}
When the EC in Protocol~\ref{alg-feedback} is taken to be a Quasi-Ideal Clock then, for any $\nu>0$ and for large enough~$d$,
\begin{align}\label{out-bound-quasiideal-feedback}
\Sigma^{{\rm out}}(\epsilon_0) <\frac{2\Sigma^{{\rm in}}(\epsilon)}{d^{1-\nu}}.
\end{align}
\end{cor}

\section{Incoherent protocols}\label{sec-incoherent}

We have seen that Protocols~\ref{alg-nofeedback} and~\ref{alg-feedback} fare well in the task of enhancing the accuracy of a clock and that their working relies heavily on the usage of a quantum enhancing clock (EC), one that satisfies the stability criterion (see Definition \ref{defi-clock}). A natural question is whether the EC being quantum is necessary for the stability. Could there be a classical EC whose error is independent of the starting position of its state? We proceed to argue that this is impossible. Nevertheless, we provide examples of protocols that work for classical ECs and also lead to an accuracy enhancement, but which is significantly lower than that achieved by Protocols~\ref{alg-nofeedback} and~\ref{alg-feedback} with quantum ECs.

Firstly, note that it is necessary that the EC evolves without any dissipation when its detector is switched off. If this is not the case, the EC will always accumulate some error, even if it is in the no ticking dynamics $\map{D}_{\rm no-tick}$. Following Protocols~\ref{alg-nofeedback} and~\ref{alg-feedback}, the dissipation will add to $\sigma^{{\rm out}}$ an additional term that depends on the starting position of the clock state.



With this requirement in mind, we can argue that the EC, as long as it is finite-dimensional, cannot be classical. The evolution of a classical clock is most generally described by a continuous-time Markov chain (see~\cite{woods2018quantum} for a detailed description of a classical clock). If it has a period of $T$ for a state $A$, we can reduce the Markov chain to a two-dimensional space spanned by $A$ and $\bar{A}$ (the other non-$A$ states). A two-state continuous-time Markov chain has a transition matrix of the form 
\begin{align*}
 P(t)=\left(\begin{array}{cc} \frac{\beta}{\alpha+\beta}+\frac{\alpha}{\alpha+\beta}e^{-t(\alpha+\beta)} & \frac{\alpha}{\alpha+\beta}-\frac{\alpha}{\alpha+\beta}e^{-t(\alpha+\beta)} \\ \frac{\beta}{\alpha+\beta}-\frac{\beta}{\alpha+\beta}e^{-t(\alpha+\beta)} & \frac{\alpha}{\alpha+\beta}+\frac{\beta}{\alpha+\beta}e^{-t(\alpha+\beta)}\end{array}\right)
 \end{align*}
with non-negative $\alpha$ and $\beta$. A periodic behaviour, i.e., $A\,P(T)=A$, can thus  occur only if $\beta=1$ and $\alpha=0$. In this case, $P(t)=I$,  i.e., $A$ is a stationary state that doesn't evolve at all. This shows that a non-stationary classical clock cannot be periodic, and is also dissipative in general. In fact, since such a clock is always fully incoherent and thus can be measured without being disturbed, turning off its detector does not reduce its inaccuracy at all. This shows that it is crucial for the EC to be a quantum clock, when considering any dynamics-switching protocol like Protocols~\ref{alg-nofeedback} and~\ref{alg-feedback}. 
 
Protocols~\ref{alg-nofeedback} and~\ref{alg-feedback} make use of an interactive control of the dynamics of the EC, which is switched between  the two modes $\map{D}_{\rm no-tick}$ and $\map{D}_{\rm tick}$ depending on the input signal. One may then ask whether this interactive switching is necessary for accuracy enhancement. In the following we consider alternative protocols that do not use such a switching of the dynamics. While we do not have a general bound, we present two natural protocols of this type and show that they achieve accuracy enhancement, although by a lower factor. For simplicity, we focus on the first output tick and drop the notation for $\epsilon$.

The first of these protocols uses the EC simply as a $d$-dimensional classical memory, whose content we denote by~$c$. As we shall see, it reduces the input signal's inaccuracy by a factor of $\sqrt{d}$.

\begin{algorithm}[H]
  \caption{Accuracy enhancement by bunching the ticks of the input signal.}
  \label{alg-input}
   \begin{algorithmic}[1]
   \Statex (Initialization) Set the EC to state $c=0$;
   \Loop
   \State Wait for an input tick,
   \If {$c=d-1$,} {produce an output tick and reset the EC to the initial state $c=0$;}
   \Else  \; \; {$c\to c+1$.}
	\EndIf
	\EndLoop
   \end{algorithmic}
\end{algorithm}

The protocols basically just bunches together $d$ ticks of the input signal to one output tick.  When the input ticks are i.i.d., the output tick has an inaccuracy that is proportional to $\Sigma^{{\rm in}}/\sqrt{d}$, because the time in between ticks is increased by a factor of $d$ whereas the confidence interval grows by a factor of $\sqrt{d}$. The output signal is then also i.i.d.  

Compared to Protocol~\ref{alg-nofeedback}, the inaccuracy reduction of Protocol~\ref{alg-input} is lower by a factor of~$\sqrt{d}$. Moreover, the latter achieves a \emph{pure} enhancement, in the sense that only the width of the tick's confidence interval is reduced. Conversely,  Protocol~\ref{alg-input} increases both the time between ticks and the width of the confidence interval. It thus does not conserve the frequency of the input signal, so that, depending on the application,  the output signal may be less useful.

\begin{figure}[b!]
	\centering
	\includegraphics[width=\linewidth]{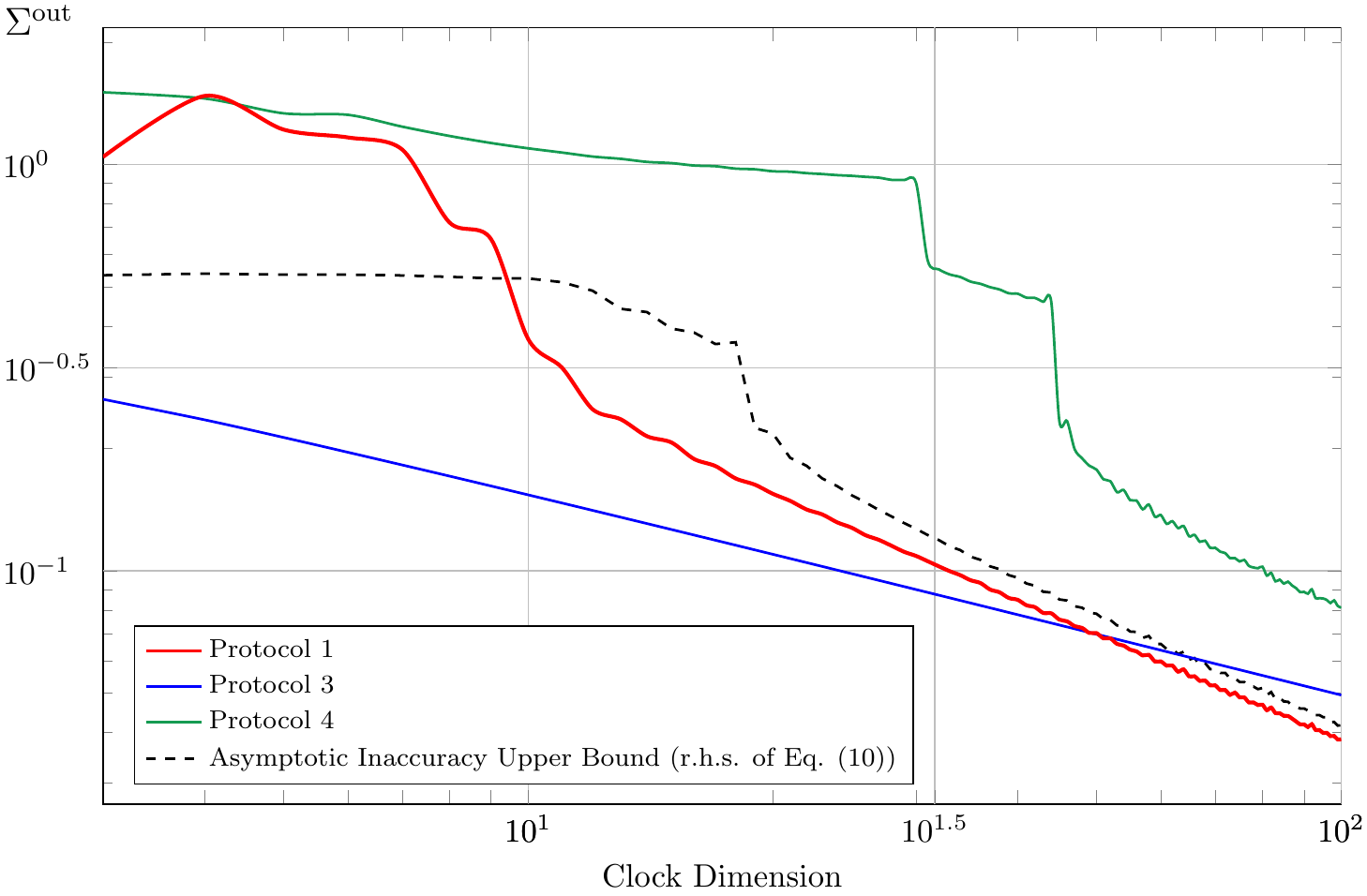}
	\caption{{\bf Numerical results for the performance of different accuracy enhancing protocols.}  The graph shows the inaccuracy~$\Sigma^{\rm out} = \Sigma^{\rm out}(\epsilon_0)$ for 	Protocols~\ref{alg-nofeedback}, \ref{alg-input}, and~\ref{alg-clock}. The input is assumed to be i.i.d.\ and box-shaped with $\Sigma^{{\rm in}}(\epsilon) \approx 0.33$ and $\epsilon = 0.01$. The horizontal axis shows the dimension~$d$ of the EC (or the classical memory) in logarithmic scale. The confidence on the output is chosen to be the same as the one on the input, i.e.\ $\epsilon_0 = 0.01$.  For both the dynamic switching protocol and the EC-tick bunching protocol the scaling is approximately $d^{-1}$, whereas for the input-tick bunching protocol it is $\smash{d^{-\frac12}}$. The black dotted line corresponds to the asymptotic upper bound on the output inaccuracy, Eq.~\eqref{out-bound-general} (evaluated for an EC with tail probability $\epsilon^{{\rm EC}_d} = 0.001$), which is seen to be quite tight already above $d \approx 20$ .}
	\label{fig_Coh_Incoh}
\end{figure}

Finally, we consider another tick-processing protocol, which is capable of preserving the frequency of the input signal. Imagine that we are given an EC with confidence interval $(\mu^{{\rm EC}_d}-\sigma^{{\rm EC}_d}/2,\mu^{{\rm EC}_d}+\sigma^{{\rm EC}_d}/2)$ where $\mu^{{\rm EC}_d}$ and $\sigma^{{\rm EC}_d}$ are much smaller than those of the input clock. This protocol ignores all ticks of the EC except the ones that follow immediately after an input tick.

\begin{algorithm}[H]
  \caption{Accuracy enhancement without controlling the dynamics of the enhancing clock.}
  \label{alg-clock}
   \begin{algorithmic}[1]
   \Loop
   \State Wait for an input tick.
   \State Wait for an EC tick.
   \State Produce an output tick.
   \EndLoop
   \end{algorithmic}
\end{algorithm} 

The inaccuracy of this protocol can be evaluated as follows. If the input tick's confidence interval is narrower than the EC's period, $\sigma^{{\rm in}} < \mu^{{\rm EC}_d}$, the number of ignored ticks can be made a constant.
Then the output's inaccuracy equals the inaccuracy of $j$ ticks of the EC bunched together, where $j\propto\mu^{{\rm in}}/\mu^{{\rm EC}_d}$ is the number of ignored ticks.
Notice further that the output mean is close to the input mean. By Eq.\ (\ref{iid-accuracy}), the output tick has inaccuracy
\begin{align}
\Sigma^{{\rm out}}&\approx \sqrt{\frac{\mu^{{\rm EC}_d}}{\mu^{{\rm in}}}}\cdot \Sigma^{{\rm EC}_d}\nonumber\\
&=\sqrt{\Sigma^{{\rm in}}}\cdot \Sigma^{{\rm EC}_d}\cdot\sqrt{\frac{\mu^{{\rm EC}_d}}{\sigma^{{\rm in}}}}\nonumber\\
&\ge \sqrt{\Sigma^{{\rm in}}}\cdot \Sigma^{{\rm EC}_d}.
\end{align} 
We may again use the Quasi-Ideal Clock for the EC, in which case the lower bound on the inaccuracy takes the form $ \sqrt{\Sigma^{{\rm in}}}/d^{1-\nu}$. 

 \begin{table}[tb]
\begin{center}
    \begin{tabular}{ | c | c | c |}
    \hline
  Protocols & Inaccuracy scaling & Rel.\ frequency\\ \hline
Dynamic switching (\ref{alg-nofeedback} \& \ref{alg-feedback}) & $\Sigma^{{\rm in}} / d^{1-\nu}$ &  $\approx 1$\\  
Bunching  input ticks (\ref{alg-input}) & $\Sigma^{{\rm in}}/  \sqrt{d}$ & $\propto d^{-1}$ \\  
Bunching  EC ticks (\ref{alg-clock}) & $\sqrt{\Sigma^{{\rm in}}} / d $ & $\approx 1$\\ \hline
\end{tabular}\caption{{\bf Asymptotic performance of the different accuracy enhancing protocols.} The table shows how the inaccuracy of the first tick scales with the inaccuracy $\Sigma^{\rm in}$ of the input signal and the dimension~$d$ of the EC, which for concreteness we take to be the Quasi-Ideal Clock. The results hold for any $\nu>0$ and sufficiently large~$d$, up to a multiplicative constant that depends on the chosen confidence level. The last column shows the relative frequency between the output and the input signal. }\label{table-compare}
\end{center}
\end{table}

From this analysis, we can see that this protocol is also outperformed by the dynamics-switching protocol (Protocol~\ref{alg-nofeedback}), since it produces redundant internal clock ticks, and by bunching together these ticks one adds a factor proportional to $\sqrt{\mu^{{\rm in}}/\mu^{{\rm EC}_d}}$ to the width of the confidence interval. 
In summary, both tick-processing protocols lose some accuracy compared to the dynamics-switching protocol, which is made apparent in Table \ref{table-compare}. A numerical simulation, depicted in Fig. \ref{fig_Coh_Incoh}, shows that the protocols of bunching ticks and the protocol of switching dynamics are all capable of achieving an enhancement, while the performance of the latter is clearly above that of the other two protocols, once the size of the clock state space is large enough.
Finally, we remark that these protocols also work with feedback, analogously to Protocol~\ref{alg-feedback}.

\section{Establish a shared clock signal}\label{sec-network}

In this section, we show that our protocols can be employed to establish a shared clock signal for multiple, and possibly distant, nodes in a network, which is crucial for various applications~\cite{lamport1978time,mills1991internet,elson2001time,elson2002fine}.

\begin{figure}[t!]
\begin{center}
  \includegraphics[width=\linewidth]{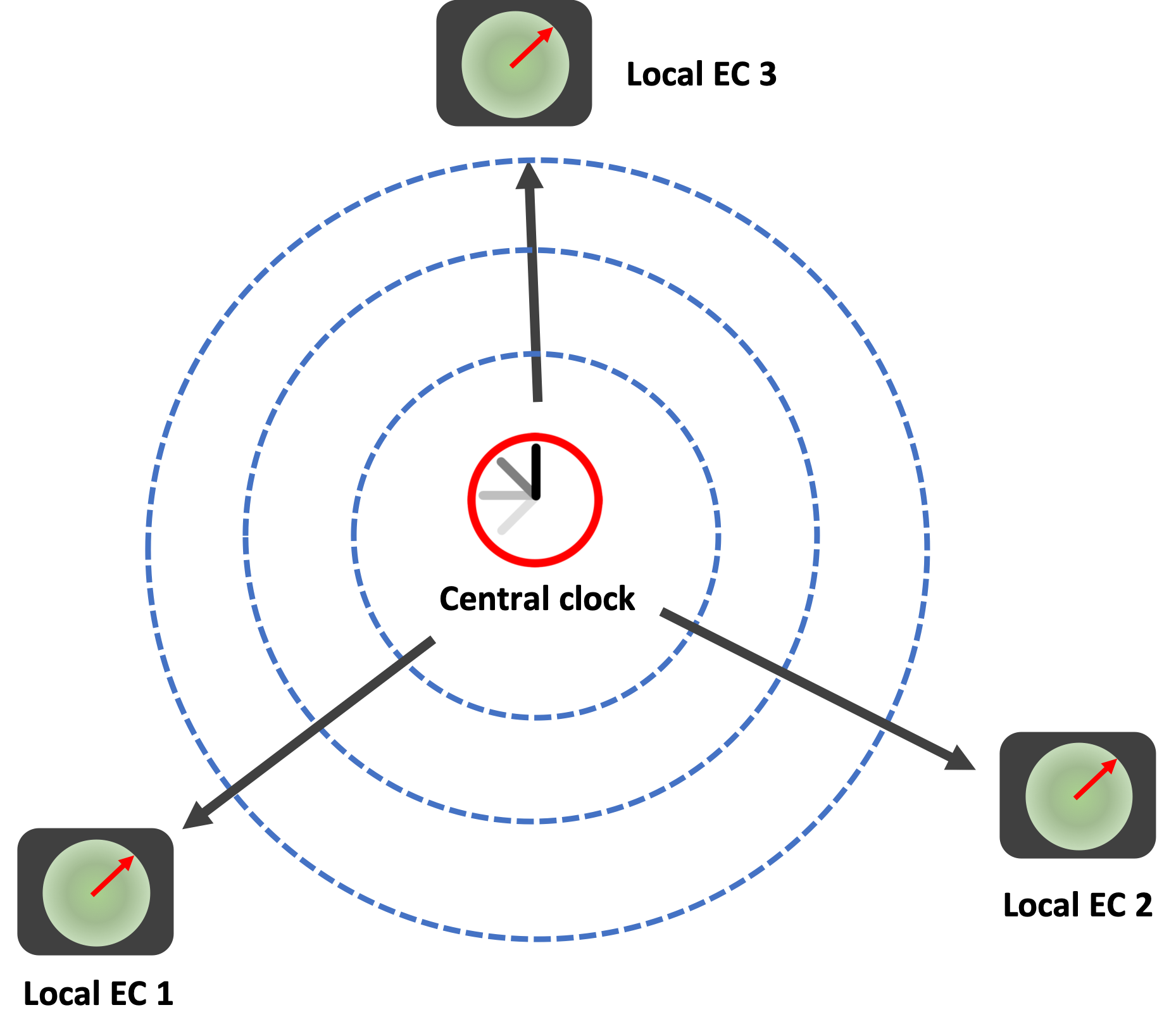}
  \end{center}
  \caption{{\bf Establish a shared clock signal.} A central clock broadcasts a clock signal (with the blue dashed circle representing the broadcast ticks) through a network of multiple nodes, in order to establish a shared time reference among them. Slight variations in the distances to the central clock lead to an error in the shared signal. Such an error can be reduced by using our accuracy enhancing protocol (Protocol \ref{alg-nofeedback}) with the pre-synchronized ECs.
  }\label{fig:network}
\end{figure}

Consider a network of multiple nodes, each having its own local clock (assumed to be a finite dimensional quantum system), as illustrated in Fig.\ \ref{fig:network}. The goal is to establish a shared signal that is accessible to all nodes in the network, in the sense that every node should hear the ticks at nearly the same time.
In order to establish the shared signal, the nodes can either synchronize the frequency and the phase of their local clocks at the beginning and refer to these local clocks afterwards, or have a central clock broadcast a clock signal through the whole network. These two obvious means of establishing a shared signal, however, are subject to various types of errors. For instance, the local signals drift away from each other rather quickly, and the broadcast signal usually has different times of arrival for different nodes (as in Fig.\ \ref{fig:network}).

Our accuracy enhancing protocols can be readily employed and outperform the obvious means in this task, by combining the broadcast signal with each of the local signals. Explicitly, the nodes would have synchronized quantum clocks at the beginning, with detectors switched off. When there is a need of establishing a shared signal, each node can locally run the enhancing protocols with the central clock being the input clock and its local clock being the EC, leading to an accuracy enhancement. In particular, if the coherent protocol (Protocol \ref{alg-nofeedback}) is employed, the inaccuracy can be lowered down to the product of the inaccuracies of the two obvious protocols.

Our protocol does not require any entanglement to be shared among the nodes, and is thus complementary to existing proposals to enhance the synchronisation of clock networks via quantum correlations \cite{komar2014quantum}. This is especially useful if a shared synchronised signal is itself necessary for measurements on entangled states (for example, given a continuous source of entangled pairs of systems).

\section{Discussion and Conclusion}\label{sec-discussion}

In this work, we proposed protocols for enhancing the accuracy of an input signal using an enhancing clock (EC). With a quantum EC of dimension~$d$, Protocols~\ref{alg-nofeedback} and~\ref{alg-feedback} achieve a reduction of the inaccuracy by a factor~$d$. The efficacy of the protocols depend crucially on the quantum nature of the EC and the possibility to adaptively control the dynamics of the EC.

Besides the application that we discussed in the network setting, our protocols can also be used as a subroutine of a highly accurate clock comprised of a macroscopic oscillator producing clock signals and a quantum system that further improves their accuracy.

Accuracy enhancing can be regarded as a signal processing task~\cite{Oppenheim1996signals}, where an input signal is processed by a special-purpose system. A key difference of clock signal processing from general signal processing is that no time reference other than the input signal is available. Common operations in signal processing, like time shifts, are therefore prohibited, which makes the task harder. Our work represents a first concrete step towards a theory of clock signal processing by harnessing properties of quantum devices.

\begin{acknowledgements}
We thank Mischa Woods for comments and technical discussions regarding the Quasi-Ideal Clock. This work is supported by the Swiss National Science Foundation via the National Center for Competence in Research ``QSIT" as well as via project No.\ 200020\_165843. 
\end{acknowledgements}

\bibliography{ref}
\begin{widetext}
\appendix

\section{Proof of Theorem \ref{theo-main}.}\label{sec-proof-nofeedback}
\subsection{A quantum clock with a switch}
An autonomous clock \cite{woods2018quantum,erker2014quantum,rankovic2015quantum,erker2017autonomous,woods2018autonomous} is identified by a tuple $\left(\rho^{{\rm C}},\{\map{M}_\delta^{{\rm C\to CT}}\}_{\delta}\right)$, where $\rho^{{\rm C}}$ is the clock state and $\{\map{M}^{{\rm C\to CT}}_\delta:\Lin(\spc{H}^{\rm C})\to\Lin(\spc{H}^{\rm C}\otimes\spc{H}^{\rm T})\}_{\delta}$ is a family of completely positive trace-preserving maps determining the evolution of the clock and the production of ticks for a duration $\delta$. Here $\spc{H}^{\rm C}$ is the Hilbert space of the clock state and $\spc{H}^{\rm T}$ is the tick space. The map $\map{M}_\delta^{{\rm C\to CT}}$ should also satisfy a continuity condition $\lim_{\Delta\to0}\lim_{\delta\to0}(\Tr_{\rm T}\circ\map{M}_\delta^{{\rm C\to CT}})^{\lfloor\Delta/\delta\rfloor}=\map{I}^{\rm C}$.
As its name suggests, an autonomous clock produces ticks without any additional input signal as reference.
An example of autonomous clocks that will be useful in our work is the Quasi-Ideal Clock, and more details of the autonomous clock model can be found in \cite{woods2018quantum}.

To make use of the input signal available in our setting, we need to extend the model of autonomous clocks.
For this purpose, we introduce the structure of a quantum clock with a switch, which serves as a key ingredient of our protocol. For a switch-controlled clock, the state space can be factorized as $\spc{H}^{\rm C}=\spc{H}^{\rm B}\otimes\spc{H}$, and the corresponding clock state is of the form $b\otimes\rho$, where $\rho$ is the clock state of an autonomous clock and $b\in\{|0\>\<0|, |1\>\<1|\}$ is a control bit. The dynamics map $\map{M}_\delta^{{\rm C\to CT}}$ is of the control form 
\begin{align}
\map{M}_\delta^{{\rm C\to CT}}=|0\>\<0|\otimes \map{D}_{{\rm no-tick},\delta}+|1\>\<1|\otimes \map{D}_{{\rm tick},\delta}
\end{align}
where $\map{D}_{{\rm no-tick},\delta}$ is infinitesimal form of the clock's unitary dynamics: $\map{D}_{{\rm no-tick},\delta}(\rho)=(\rho-i\delta[H,\rho])\otimes |0\>\<0|$ with $H$ being the Hamiltonian on $\spc{H}$, and $\map{D}_{{\rm tick},\delta}$ is the dynamics of an autonomous clock.
From the above description we can see that the tick production is \emph{switched off} and the clock evolves unitarily if the control bit is zero while the tick production is \emph{switched on} and the clock runs in the same pattern as the autonomous clock if the control bit is one. 
We denote by $\tau^{{\rm EC}_d}$ the period of the unitary dynamics of this autonomous clock and by $\Psi_{s}$ its initial state with $s\in(-\tau^{{\rm EC}_d}/2,\tau^{{\rm EC}_d}/2]$. By definition, if the state starts in $\Psi_{s}$ and evolves unitarily for $\tau^{{\rm EC}_d}$ it will end up in the original state $\Psi_{s}$. 
In an autonomous clock, the clock starts with its state in the initial state $\Psi_{0}$. The time it takes to tick is a random variable, denoted here as $T^{{\rm EC}_d}_s$.  When the clock has high enough dimension, $T^{{\rm EC}_d}_s$ is close to $\tau^{{\rm EC}_d}/2$ with high probability. After producing a tick, the  clock will be reset to its original state $\Psi_0$ \cite{woods2018quantum}. 

In the setting of our protocol there are input ticks that offer additional time reference to the clock. An input tick will trigger an operation on the control bit. In general, this operation could be any quantum channel $\map{C}^{{\rm b\to b}}$. In this work, we only consider simple logic operations, which will switch on/off the production of the autonomous clock's ticks. 
For instance, in our protocols, an input tick aways triggers the detector to be on, which corresponds to be operation $b\to 1$ for $b\in\{0,1\}$.
Notice that the implementation time of these operations is assumed to be zero (which is otherwise arbitrary).

\subsection{Conventions and notions}

For a fixed $\epsilon>0$, we consider an input signal with i.i.d.\ input ticks and a confidence interval
$\left(\mu^{{\rm in}}(\epsilon)-\sigma^{{\rm in}}(\epsilon)/2,\mu^{{\rm in}}(\epsilon)+\sigma^{{\rm in}}(\epsilon)/2\right)$ satisfying
\begin{align}\label{req-input}
\sigma^{{\rm in}}(\epsilon)<\frac{2\mu^{{\rm in}}(\epsilon)}{3}.
\end{align}
We denote by $\Sigma^{{\rm in}}(\epsilon):=\sigma^{{\rm in}}(\epsilon)/\mu^{{\rm in}}(\epsilon)$ its inaccuracy.
For conciseness, we shall abbreviate $\Sigma^{{\rm in}}(\epsilon)$ to $\Sigma^{{\rm in}}$ (and similarly for other $\epsilon$-dependent quantities) when there is no risk of confusion.

Before going into the proof, it is convenient to define a couple of variables that we are going to encounter frequently.
We denote by $t^{{\rm out}}_i$ the time when the $i$-th tick of the output signal is produced, and by $t^{{\rm in}}_i$ the time  when the first input tick after the $(i-1)$-th output tick arrives.
Notice that, by this definition, $t^{{\rm in}}_i-t^{{\rm in}}_{i-1}$ is no longer i.i.d.\ since there may be multiple input ticks in between.
 In particular, we define $t^{{\rm in}}_0$ as the time of arrival for the 0-th input tick. The purpose of this definition is that there may be multiple input ticks that arrive in between two consecutive output ticks but only the first one of them triggers a non-trivial operation on the clock. 
The time between $i$-th input tick and $(i-1)$-th input tick is a random variable, denoted as $T^{{\rm in}}_i$. 

For convenience of analysis, we choose $\tau^{{\rm EC}_d}$ so that 
\begin{align}\label{mu-in}
\mu^{{\rm in}}=(m+1/2)\cdot \tau^{{\rm EC}_d}
\end{align}
 with $m\in\N^*$. The choice of $m$ will affect the performance of the protocol and will be discussed later.  For readers' convenience, notations that appear frequently in the next subsection are listed in Table \ref{table-notation}.

 \begin{table}[h!]
\begin{center}
    \begin{tabular}{ | c | c |}
    \hline
  Notation & Definition \\ \hline
$t^{{\rm out}}_i$ & the time when the $i$-th output tick is produced \\ \hline 
$t^{{\rm in}}_0$ & the arrival time of the first input tick \\ \hline 
$t^{{\rm in}}_i$ ($i\ge 1$) & the arrival time of the first input tick after the $(i-1)$-th output tick \\ \hline 
$s_i$ & the clock parameter of the quantum clock at time $t^{{\rm in}}_{i}$ \\ \hline
$T^{{\rm EC}_d}_{s_i}$ & $t^{{\rm out}}_i-t^{{\rm in}}_i$ (which depends on the state $\Psi_{s_i}$)\\ \hline
$T^{{\rm in}}$ & the time in between two consecutive input ticks \\ \hline
$\set{C}^{x}_{i}$ $(x={\rm in},{\rm out})$ & the confidence interval of $t^{x}_i$  \\ \hline
$\sigma^{{\rm in}}$ &  width of the confidence interval of input ticks \\ \hline
$\mu^{{\rm in}}$ & center of the input confidence interval $\set{C}^{{\rm in}}$ \\ \hline
$\tau^{{\rm EC}_d}$ & period of unitary evolution of the quantum clock ($\tau^{{\rm EC}_d}\approx 2\mu^{{\rm EC}_d}$) \\ \hline
\end{tabular}\caption{Overview on the notation.}\label{table-notation}
\end{center}
\end{table}

\subsection{Confidence intervals of the output ticks}
 
From this subsection, we start to bound the accuracy of the first $j\in\N^*$ ticks, which requires us to determine a confidence interval of the random variable  $t^{{\rm out}}_{j}-t^{{\rm out}}_{0}$. 
The key step of the proof is to show that all the input ticks and the output ticks, characterized by the random variables $t^{{\rm in}}_{i}$ and $t^{{\rm out}}_{i}$ ($i\le j$), fall within narrow confidence intervals around their expectations with large probability. In this case, the output ticks are generated \emph{as expected}. In the case where either an input tick or an output tick falls outside its respective confidence interval, the output ticks may be triggered too early or too late, resulting in a large error. This idea is made precise by the following lemma:
\begin{lemma}\label{lem-confidence}
Assume for convenience $t^{{\rm in}}_{0}=0$, which is otherwise arbitrary.
For $j\in\N^*$ such that 
\begin{align}\label{cond-j}
j<\frac{\tau^{{\rm EC}_d}-\sigma^{{\rm EC}_d}}{\sigma^{{\rm EC}_d}+\sigma^{{\rm in}}},
\end{align}
 the probability that $t^{{\rm in}}_{i}\in\set{C}^{{\rm in}}_{i}$ and $t^{{\rm out}}_{i}\in\set{C}^{{\rm out}}_{i}$ holds for every $i\le j$ is at least $1-\epsilon^{{\rm out}}_j$, where the confidence intervals for the input tick and the output tick are defined as
\begin{align}\label{conf-T-in}
\set{C}^{{\rm in}}_i:=i\cdot\left(\mu^{{\rm in}}-\frac{\sigma^{{\rm in}}}2,\mu^{{\rm in}}+\frac{\sigma^{{\rm in}}}2\right)
\end{align}
and 
\begin{align}\label{conf-T-out}
\set{C}^{{\rm out}}_i:=\left\{\begin{array}{ll}\left(\frac{\tau^{{\rm EC}_d}-\sigma^{{\rm EC}_d}}{2},\frac{\tau^{{\rm EC}_d}+\sigma^{{\rm EC}_d}}{2}\right)&\qquad i=0\\
i\cdot\left(\mu^{{\rm in}}-\frac{\sigma^{{\rm EC}_d}}2,\mu^{{\rm in}}+\frac{\sigma^{{\rm EC}_d}}2\right)+t^{{\rm out}}_{0}&\qquad i\ge 1.\\
\end{array}\right.
\end{align}
Here $a\cdot(b,c)+d$ is a shorthand for $(ab+d,ac+d)$. 
The tail probability of the output confidence interval is bounded as
\begin{align}\label{epsilon-j}
\epsilon^{{\rm out}}_j\le  j\cdot\epsilon+(j+1)\cdot\epsilon^{{\rm EC}_d}
\end{align}
where $\epsilon$ is the tail probability of the input ticks and $\epsilon^{{\rm EC}_d}$ is the tail probability of the quantum clock that vanishes as $d\to\infty$. 

\end{lemma}

It is straightforward to check that the conditions (\ref{mu-in}) and (\ref{cond-j}) guarantee that all the confidence intervals have no intersection and are temporally ordered.
If all input and output ticks are in their own confidence intervals, proper causal order among them will be ensured, i.e.
\begin{align}\label{temp}
t^{{\rm in}}_{0}\le t^{{\rm out}}_{0}\le t^{{\rm in}}_{1} \le\cdots\le t^{{\rm out}}_{j-1}\le t^{{\rm in}}_{j}\le t^{{\rm out}}_{j}.
\end{align}
From Eq.\ (\ref{conf-T-out}) we can see that the width of an output tick's confidence interval is $\sigma^{{\rm EC}_d}$, which is a quantity dependent on the quantum clock and independent of the input tick. 
This feature is key to the accuracy enhancement, which is a result of the stability criterion.

\noindent{\em Proof of Lemma \ref{lem-confidence}.}
Applying the chain rule, we can express the probability that all ticks are in their confidence intervals as
\begin{align}
\Pr\left[\bigcap_{i=0}^{j}\left(t^{{\rm in}}_{i}\in\set{C}^{{\rm in}}_{i}\cap t^{{\rm out}}_{i}\in\set{C}^{{\rm out}}_{i}\right)\right]&=\prod_{i=0}^{j}\left(\Pr\left[t^{{\rm in}}_{i}\in\set{C}^{{\rm in}}_{i}\,\big|\,\bigcap_{l=1}^{i-1}\left(t^{{\rm in}}_{l}\in\set{C}^{{\rm in}}_{l}\cap t^{{\rm out}}_{l}\in\set{C}^{{\rm out}}_{l}\right)\right]\right.\label{chain-rule}\\
&\qquad\left.\times\Pr\left[t^{{\rm out}}_{i}\in\set{C}^{{\rm out}}_{i}\,\big|\,\bigcap_{l=1}^{i-1}\left(t^{{\rm in}}_{l}\in\set{C}^{{\rm in}}_{l}\cap t^{{\rm out}}_{l}\in\set{C}^{{\rm out}}_{l}\right)\cap t^{{\rm in}}_{i}\in\set{C}^{{\rm in}}_{i}\right]\right).\nonumber
\end{align}
First we bound the probability for $t^{{\rm in}}_{i}$ to be in its confidence interval given that all previous ticks are within their confidence intervals.
Notice that $t^{{\rm in}}_{0}=0\in\set{C}^{{\rm in}}_{1}$ trivially holds. For $i\ge1$, we stress that the probability of $t^{{\rm in}}_{i}\in\set{C}^{{\rm in}}_{i}$ does not simply follow from the i.i.d. property of the input ticks, because there may be other input ticks in between $t^{{\rm out}}_{i-1}$ and $t^{{\rm in}}_{i-1}\in\set{C}^{{\rm in}}_{i-1}$.
Instead, consider the time of arrival of the next input tick after $t^{{\rm in}}_{i-1}$, conditioned on $t^{{\rm out}}_{i-1}\in\set{C}^{{\rm out}}_{i-1}$ and $t^{{\rm in}}_{i-1}\in\set{C}^{{\rm in}}_{i-1}$. 
Since $\mu^{{\rm in}}-\sigma^{{\rm in}}/2>\tau^{{\rm EC}_d}>t^{{\rm out}}_{i-1}-t^{{\rm in}}_{i-1}$ [see Eq. (\ref{mu-in}) and Eq. (\ref{cond-j})],
if the time it takes this tick to arrive is $T^{{\rm in}}\in(\mu^{{\rm in}}-\sigma^{{\rm in}}/2,\mu^{{\rm in}}+\sigma^{{\rm in}}/2)$, then it is clear that this tick will be the first input tick after $t^{{\rm out}}_{i-1}$. In formula, this argument reads
\begin{align}\label{tailprob-in}
\Pr\left[t^{{\rm in}}_{i}\in\set{C}^{{\rm in}}_{i}\,\big|\,\bigcap_{l=0}^{i-1}\left(t^{{\rm in}}_{l}\in\set{C}^{{\rm in}}_{l}\cap t^{{\rm out}}_{l}\in\set{C}^{{\rm out}}_{l}\right)\right]\ge 1-\epsilon.
\end{align}

Now we bound the probability for $t^{{\rm out}}_{i}$ to be in their confidence intervals, conditioned on all the previous ticks are in their confidence intervals.
Define $s_i\in(-\tau^{{\rm EC}_d}/2,\tau^{{\rm EC}_d}/2]$ to be so that the clock state is $\Psi_{s_i}$ at $t_i^{{\rm in}}$. Intuitively, it is the location of the  clock's ``hand'' at time $t_i^{{\rm in}}$. 
Since the clock is in the initial state $\Psi_0$ at time $t^{{\rm out}}_{i-1}$, we obtain the following relation
\begin{align}\label{cond-mj}
t^{{\rm in}}_i-t^{{\rm out}}_{i-1}=s_i+m_i\cdot\tau^{{\rm EC}_d}
\end{align}
where $m_i\in\N$ is the number of periods that the clock has been evolving unitarily.
Notice that when $t^{{\rm in}}_{i}\in\set{C}^{{\rm in}}_{i}$ and $t^{{\rm out}}_{i-1}\in\set{C}^{{\rm out}}_{i-1}$, we have
\begin{align}\label{cond-mj2}
s_i+m_i\cdot\tau^{{\rm EC}_d}\in\left(m\cdot\tau^{{\rm EC}_d}-\frac{i(\sigma^{{\rm EC}_d}+\sigma^{{\rm in}})}{2},m\cdot\tau^{{\rm EC}_d}+\frac{i(\sigma^{{\rm EC}_d}+\sigma^{{\rm in}})}{2}\right).
\end{align}
Since $\frac{i(\sigma^{{\rm EC}_d}+\sigma^{{\rm in}})}{2}<\frac{\tau^{{\rm EC}_d}-\sigma^{{\rm EC}_d}}{2}$ holds thanks to Eq. (\ref{cond-j}), Eq. (\ref{cond-mj2}) implies that 
\begin{align}\label{m}
m_i=m
\end{align} 
and 
\begin{align}\label{cond-tistar}
s_i\in\left(-\frac{\tau^{{\rm EC}_d}-\sigma^{{\rm EC}_d}}{2},\frac{\tau^{{\rm EC}_d}-\sigma^{{\rm EC}_d}}{2}\right).
\end{align}

Next, we denote by $T^{{\rm EC}_d}_{s_i}:=t^{{\rm out}}_i-t^{{\rm in}}_i$ the time that the clock evolves non-unitarily in between two output ticks, whose value can be determined by $s_i$ and the stability criterion.
Bringing together Eq. (\ref{cond-mj}), Eq. (\ref{m}) and the definition of $T^{{\rm EC}_d}_{s_i}$, we have 
\begin{align}
t^{{\rm out}}_{i}&=t^{{\rm out}}_{i-1}+m\cdot\tau^{{\rm EC}_d}+\left(T^{{\rm EC}_d}_{s_i}+s_{i}\right).
\end{align}
Noticing that the stability criterion is guaranteed by Eq. (\ref{cond-tistar}), we can apply the stability criterion to $T^{{\rm EC}_d}_{s_i}+s_{i}$, which yields that
\begin{align}
&\Pr\left[t^{{\rm out}}_{i}-t^{{\rm out}}_{i-1}-m\cdot\tau^{{\rm EC}_d}\in\left(\frac{\tau^{{\rm EC}_d}-\sigma^{{\rm EC}_d}}{2},\frac{\tau^{{\rm EC}_d}+\sigma^{{\rm EC}_d}}{2}\right)\,\big|\,\bigcap_{l=0}^{i-1}\left(t^{{\rm in}}_{l}\in\set{C}^{{\rm in}}_{l}\cap t^{{\rm out}}_{l}\in\set{C}^{{\rm out}}_{l}\right)\cap t^{{\rm in}}_{i}\in\set{C}^{{\rm in}}_{i}\right]\ge 1-\epsilon^{{\rm EC}_d}.
\end{align}
Since $t^{{\rm out}}_{i}-t^{{\rm out}}_{i-1}-m\cdot\tau^{{\rm EC}_d}\in\left(\frac{\tau^{{\rm EC}_d}-\sigma^{{\rm EC}_d}}{2},\frac{\tau^{{\rm EC}_d}+\sigma^{{\rm EC}_d}}{2}\right)$ plus $t^{{\rm out}}_{i-1}\in\set{C}^{{\rm out}}_{i-1}$ imply that $t^{{\rm out}}_{i}\in\set{C}^{{\rm out}}_{i}$, we conclude that 
\begin{align}\label{tailprob-c}
\Pr\left[t^{{\rm out}}_{i}\in\set{C}^{{\rm out}}_{i}\,\big|\,\bigcap_{l=0}^{i-1}\left(t^{{\rm in}}_{l}\in\set{C}^{{\rm in}}_{l}\cap t^{{\rm out}}_{l}\in\set{C}^{{\rm out}}_{l}\right)\cap t^{{\rm in}}_{i}\in\set{C}^{{\rm in}}_{i}\right]\ge 1-\epsilon^{{\rm EC}_d}.
\end{align}

Finally, substituting Eq. (\ref{tailprob-in}) and Eq. (\ref{tailprob-c}) into Eq. (\ref{chain-rule}), we have
\begin{align}
\Pr\left[\bigcap_{i=0}^{j}\left(t^{{\rm in}}_{i}\in\set{C}^{{\rm in}}_{i}\cap t^{{\rm out}}_{i}\in\set{C}^{{\rm out}}_{i}\right)\right]\ge 1-\epsilon^{{\rm out}}_j
\end{align}
where $\epsilon^{{\rm out}}_j\le j\cdot\epsilon+(j+1)\cdot\epsilon^{{\rm EC}_d}$ as desired.

\qed

\begin{rem}
Eq. (\ref{cond-j}) puts a limit on how small we can set $\tau^{{\rm EC}_d}$ to be. When the clock system has large dimension, the term $\sigma^{{\rm EC}_d}$ is very small and Eq. (\ref{cond-j}) becomes
\begin{align}
\tau^{{\rm EC}_d}> j\cdot\sigma^{{\rm in}}.
\end{align}
Therefore, the protocol cannot run forever with small error, since the above constraint is always going to be violated when $j$ is large enough. To address this issue, one can choose to reset the output signal every once in a while.
\end{rem}

\subsection{Accuracy of the output signal}

In this subsection, we evaluate the accuracy of the output signal using Lemma \ref{lem-confidence}.
First, we emphasize that Eq. (\ref{cond-j}) in Lemma \ref{lem-confidence} holds at least for $j=1$ when the input signal satisfies the condition (\ref{req-input}), as we have $\tau^{{\rm EC}_d}>\sigma^{{\rm in}}$ by setting $m=1$ in Eq. (\ref{mu-in}).
Noticing that $\sigma^{{\rm EC}_d}$ vanishes in the large $d$ limit, we have $\tau^{{\rm EC}_d}>  \sigma^{{\rm in}}+2\sigma^{{\rm EC}_d}$ for large enough $d$, and thus Eq. (\ref{cond-j}) holds at least for $j=1$.

Now we show a lower bound of the output accuracy.
It is straightforward from Eq. (\ref{conf-T-out}) that there exists a confidence interval of the $j$-th output tick with center and width
\begin{align}
\mu^{{\rm out}}(\epsilon_j)=j\cdot\mu^{{\rm in}}\qquad \sigma^{{\rm out}}(\epsilon_j)=j\cdot\sigma^{{\rm EC}_d},
\end{align}
which has a tail probability $\epsilon^{{\rm out}}_j$ given by Eq. (\ref{epsilon-j}). Therefore, the output accuracy can be evaluated as
\begin{align}\label{out-bound-temp}
\Sigma^{{\rm out}}_{j}(\epsilon_j)= \frac{j\cdot\sigma^{{\rm EC}_d}}{\mu^{{\rm in}}} =\frac{j\cdot\tau^{{\rm EC}_d}}{2\sigma^{{\rm in}}}\cdot \Sigma^{{\rm in}}(\epsilon)\cdot \bar{\Sigma}^{{\rm EC}_d}.
\end{align}
Notice that for large enough $d$ Eq. (\ref{cond-j}) becomes $j\cdot\sigma^{{\rm in}}<\tau^{{\rm EC}_d}$.
Choose $m$ in Eq. (\ref{mu-in}) as large as possible so that this inequality  ``barely holds'', in the sense that
\begin{align}
j\cdot\sigma^{{\rm in}}\in\left[\frac{\mu^{{\rm in}}}{m+3/2},\frac{\mu^{{\rm in}}}{m+1/2}\right).
\end{align}
Then the ratio between $\sigma^{{\rm in}}(\epsilon)$ and $\tau^{{\rm EC}_d}/2$ can be bounded as $\frac{\sigma^{{\rm in}}}{\tau^{{\rm EC}_d}/2}\ge \frac{2(m+1/2)}{j(m+3/2)}\ge \frac{6}{5j}$.
Substituting it into Eq. (\ref{out-bound-temp}), we get the bound (\ref{out-bound-general}).

For the Quasi-Ideal Clock, the inaccuracy is at most $\bar{\Sigma}^{{\rm EC}_d}=2d^{-1+\eta}\left(1+O(d^{-\eta})\right)$, which is an immediate consequence of the following lemma (see Appendix \ref{sec-lemma1} for its proof):
\begin{lemma}[Quasi-Ideal Clock with arbitrary initial position]\label{lemma-quasiideal}
A $d$-dimensional Quasi-Ideal Clock satisfies the stability criterion. Moreover, the width of the confidence interval is
\begin{align}\label{sigma-clock}
\sigma^{{\rm EC}_d}:=\left(\gamma+\frac{x_{\rm vr}}{\pi}\right)\tau^{{\rm EC}_d}.
\end{align}
Here $\gamma=d^{-1+\eta}+O(d^{-1})$ for any $\eta>0$ \cite[Eqs. (F23) and (F24)]{woods2018quantum} and $x_{\rm vr}=(1/\pi)d^{\frac{3\eta}{4}-1}$ \cite[Eq. (F22)]{woods2018quantum}, thus the leading order term in Eq. (\ref{sigma-clock}) is $\gamma\cdot\tau^{{\rm EC}_d}$.
The tail probability is
\begin{align}\label{error-clock}
\epsilon^{{\rm EC}_d}=2\delta\tilde{\epsilon}_V+e^{-\delta}+3\epsilon_{\rm tail}+2\epsilon_{\rm trans}\left(\tau^{{\rm EC}_d}\right).
\end{align}
The major term in Eq. (\ref{error-clock}) is $2\delta\tilde{\epsilon}_V=o(\gamma)$ (see \cite[Corollary 9 and Eq. (F240)]{woods2018quantum}), whereas $\delta=d^{\frac{\eta}{16}}$ (see \cite[Eq. (F19)]{woods2018quantum}), the other two overhead terms $\epsilon_{\rm tail}$ and $\epsilon_{\rm trans}$ also vanish exponentially in $d$ and are given in Section \ref{sec-lemma1}.  
\end{lemma}
Substituting the expression of $\bar{\Sigma}^{{\rm EC}_d}$ for the Quasi-Ideal Clock into Eq. (\ref{out-bound-general}), we get Eq. (\ref{out-bound-quasiideal}).

\section{Proof of Theorem \ref{theo-feedback}}\label{sec-proof-feedback}
Here we prove Theorem \ref{theo-feedback} on the accuracy of the output signal for the protocol with feedback. The proof is similar to the proof of Theorem \ref{theo-main} but essentially simpler thanks to the reset mechanism. First, since the output ticks of the quantum clock are i.i.d., we only need to evaluate the accuracy for $t:=t^{{\rm out}}_{1}-t^{{\rm out}}_{0}$ and the accuracy of the $j$-th tick can be estimated from Eq.\ (\ref{iid-accuracy}).

We choose $\tau^{{\rm EC}_d}$ so that 
\begin{align}\label{mu-in-feedback}
\mu^{{\rm in}}=m\cdot \tau^{{\rm EC}_d}
\end{align}
 with $m\in\N^*$. 
The key step of the proof is again to show that probability that the second input tick and the second output tick, characterized by the random variables $t^{{\rm in}}_{1}$ and $t^{{\rm out}}_{1}$, falls within certain intervals around their expectations with large probability. 

\subsection{Confidence intervals of the output tick}
In this subsection, we show the following lemma:
\begin{lemma}\label{lem-confidence-feedback}
Assume for convenience $t^{{\rm out}}_{0}=0$, which is otherwise arbitrary.
For a quantum clock satisfying the stability criterion in Definition \ref{defi-clock} and the constraint
\begin{align}\label{constraint-feedback}
\sigma^{{\rm in}}<\tau^{{\rm EC}_d}-\sigma^{{\rm EC}_d},
\end{align}
the probability that $t^{{\rm in}}_{1}\in\set{C}^{{\rm in}}_{1}$ and $t^{{\rm out}}_{1}\in\set{C}^{{\rm out}}_{1}$ holds is at least $1-\epsilon^{{\rm out}}$, where the confidence intervals for the input tick and the output tick are defined as
\begin{align}\label{conf-T-in-feedback}
\set{C}^{{\rm in}}_1:=\left(\mu^{{\rm in}}-\frac{\sigma^{{\rm in}}}{2},\mu^{{\rm in}}+\frac{\sigma^{{\rm in}}}{2}\right)
\end{align}
and 
\begin{align}\label{conf-T-out-feedback}
\set{C}^{{\rm out}}_1:=\left(\mu^{{\rm in}}+\frac{\tau^{{\rm EC}_d}-\sigma^{{\rm EC}_d}}{2},\mu^{{\rm in}}+\frac{\tau^{{\rm EC}_d}+\sigma^{{\rm EC}_d}}{2}\right).
\end{align}
The tail probability is bounded as $\epsilon^{{\rm out}}\le \epsilon+\epsilon^{{\rm EC}_d}$.
\end{lemma}

\noindent{\em Proof of Lemma \ref{lem-confidence-feedback}.}
By definition, the probability that $t^{{\rm in}}_1$ in its confidence interval is just bounded as
 \begin{align}
\Pr\left[t^{{\rm in}}_1\in\set{C}^{{\rm in}}_1\right]\ge 1-\epsilon.
 \end{align}
The clock state at $t^{{\rm in}}_1$ is $\Psi_{s_1}$, where 
\begin{align}\label{t2-star}
s_1=t^{{\rm in}}_1-m_1\cdot\tau^{{\rm EC}_d}\in\left(-\frac{\tau^{{\rm EC}_d}}{2},\frac{\tau^{{\rm EC}_d}}{2}\right]
\end{align}
for some $m_1\in\N$. Under the condition $t^{{\rm in}}_1\in\set{C}^{{\rm in}}_1$ and (\ref{constraint-feedback}), we have 
\begin{align}\label{m2}
m_1=m
\end{align}
and
\begin{align}
s_1\in\left(-\frac{\tau^{{\rm EC}_d}-\sigma^{{\rm EC}_d}}{2},\frac{\tau^{{\rm EC}_d}-\sigma^{{\rm EC}_d}}{2}\right).
\end{align}
We can then apply Lemma \ref{lemma-quasiideal}, which yields that
\begin{align}\label{inter-tc+t*}
\left(T^{{\rm EC}_d}_{s_1}+s_{1}\right)\in\left(\frac{\tau^{{\rm EC}_d}-\sigma^{{\rm EC}_d}}{2},\frac{\tau^{{\rm EC}_d}+\sigma^{{\rm EC}_d}}{2}\right)
\end{align}
with probability $1-\epsilon^{{\rm EC}_d}$. Here $T^{{\rm EC}_d}_{s_1}:=t^{{\rm out}}_{1}-t^{{\rm in}}_{1}$. 
Combining Eq. (\ref{t2-star}), Eq. (\ref{m2}) with the above equation, we get 
\begin{align}
t^{{\rm out}}_{1}=m_1\cdot\tau^{{\rm EC}_d}+T^{{\rm EC}_d}_{s_1}+s_{1}&\in\set{C}^{{\rm out}}_1,
\end{align}
which means that Eq. (\ref{inter-tc+t*}) implies $t^{{\rm out}}_{1}\in\set{C}^{{\rm out}}_1$. Then we conclude that
\begin{align}
\Pr\left[t^{{\rm out}}_{1}\in\set{C}^{{\rm out}}_1\,\big|\,t^{{\rm in}}_{1}\in\set{C}^{{\rm in}}_1\right]\ge 1-\epsilon^{{\rm EC}_d}.
\end{align}

Finally, by the chain rule, we have
\begin{align}
\Pr\left[t^{{\rm in}}_1\in \set{C}^{{\rm in}}_1\cap t^{{\rm out}}_1\cap\set{C}^{{\rm out}}_1\right]&=\Pr\left[t^{{\rm out}}_1\cap\set{C}^{{\rm out}}_1\,\big|\,t^{{\rm in}}_1\in \set{C}^{{\rm in}}_1\right]\Pr\left[t^{{\rm in}}_1\in \set{C}^{{\rm in}}_1\right]\\
&\ge \left(1-\epsilon^{{\rm EC}_d}\right)(1-\epsilon)
\end{align}
and $\epsilon^{{\rm out}}\le \epsilon+\epsilon^{{\rm EC}_d}$ as desired.
\qed

\subsection{Accuracy of the output signal}

In this subsection, we evaluate the accuracy of the output signal using Lemma \ref{lem-confidence-feedback}.
First, we emphasis that Eq. (\ref{constraint-feedback}) in Lemma \ref{lem-confidence-feedback} holds since $\sigma^{{\rm in}}(\epsilon)<\mu^{{\rm in}}(\epsilon)$ by assumption and $\sigma^{{\rm EC}_d}$ vanishes as $d\to\infty$ by the stability criterion.

Then we show a lower bound of the output accuracy.
It is straightforward from Eq. (\ref{conf-T-out-feedback}) that there exists a confidence interval of the output tick with center $\mu^{{\rm in}}(\epsilon)+\tau^{{\rm EC}_d}/2$ and width $\sigma^{{\rm EC}_d}$
which has a tail probability $\epsilon^{{\rm out}}$. Therefore, the output inaccuracy can be evaluated as
\begin{align}\label{out-bound-temp-feedback}
\Sigma^{{\rm out}}\left(\epsilon^{{\rm out}}\right)=\frac{\sigma^{{\rm EC}_d}}{\mu^{{\rm in}}+\tau^{{\rm EC}_d}/2}<\left(\frac{\tau^{{\rm EC}_d}/2}{\sigma^{{\rm in}}}\right)\cdot \Sigma^{{\rm in}}(\epsilon)\cdot \bar{\Sigma}^{{\rm EC}_d}.
\end{align}
Choose $m$ in Eq. (\ref{mu-in-feedback}) as large as possible so that Eq. (\ref{constraint-feedback}) ``barely holds'', in the sense that
\begin{align}
\sigma^{{\rm in}}\in\left[\frac{\mu^{{\rm in}}}{m+1},\frac{\mu^{{\rm in}}}{m}\right).
\end{align}
Then the ratio between $\sigma^{{\rm in}}$ and $\tau^{{\rm EC}_d}/2$ can be bounded as $\frac{\sigma^{{\rm in}}}{\tau^{{\rm EC}_d}/2}\ge \frac{2m}{m+1}\ge 1$.
Substituting it into Eq. (\ref{out-bound-temp-feedback}), we get the bound (\ref{out-bound-general-feedback}).
For the Quasi-Ideal Clock, the inaccuracy is at most $\bar{\Sigma}^{{\rm EC}_d}=2d^{-1+\eta}\left(1-O(d^{-\eta})\right)$. 
Substituting the expression of $\bar{\Sigma}^{{\rm EC}_d}$ for the Quasi-Ideal Clock into Eq. (\ref{out-bound-general-feedback}), we get Eq. (\ref{out-bound-quasiideal-feedback}).

\section{Proof of Lemma \ref{lemma-quasiideal}}\label{sec-lemma1}

Define $\rho_s(t):=|\psi_t\>_s\<\psi_t|_s$ where $|\psi_t\>_s:=e^{-itH-t\delta\bar{V}_d}|\Psi_{s}\>$ is the unnormalized clock state. Here $|\Psi_{s}\>$ is the initial state of the non-unitary evolution, $H$ is the Hamiltonian, $\bar{V}_d$ is a real operator corresponds to the interaction potential that generates output ticks, and $\delta>0$ is the interaction strength. Note that the real part of the exponent causes the norm of the state to decrease so that the state is not normalized.
The advantage of using this notation is that $\Tr\left[\rho_s(t)\right]$ equals the probability that the clock evolves for time $t$ without producing any tick \cite{woods2018quantum}.

Define $I_{\pm}:=\frac{\tau^{{\rm EC}_d}}{2}-s\pm \frac{\sigma^{{\rm EC}_d}}{2}$ as the left boundary and the right boundary of the confidence interval.
The probability that the tick is generated in the confidence interval $I=[I_-,I_+]$ can be expressed as
\begin{align}\label{app-interval-I}
\Pr\left[T^{{\rm EC}_d}_s\in I\right]&:=\Tr\left[\rho(I_{-})\right]-\Tr\left[\rho(I_{+})\right].
\end{align}
The statement of Lemma \ref{lemma-quasiideal}, i.e.\ Eq.\ (\ref{tailbound-clock}), can be rephrased as
\begin{align}
\Tr\left[\rho(I_{-})\right]-\Tr\left[\rho(I_{+})\right]\ge 1-\epsilon^{{\rm EC}_d}.
\end{align}
Therefore, to show an upper bound of the tail probability, we need to derive a lower bound on $\Tr\left[\rho(I_{-})\right]$ and an upper bound on $\Tr\left[\rho(I_{+})\right]$.
For this purpose, we first introduce the following lemma, which comes immediately from Lemma 21 and Lemma 22 of Ref.\ \cite{woods2018quantum}:
\begin{lemma}[\cite{woods2018quantum}, Lemma 21 and Lemma 22]\label{lemma-error}
\begin{align}
 \Delta_{\rm EC}(t)-\epsilon_{\rm tail}-\epsilon_{\rm trans}(t)\le\Tr\left[\rho(t)\right]\le \Delta_{\rm EC}(t)+\epsilon_{\rm tail}+\epsilon_{\rm trans}(t).
\end{align}
Here
\begin{align}
\epsilon_{\rm trans}(t)&=|t|\frac{d}{\tau^{{\rm EC}_d}}\left(O\left(\frac{\sigma^3}{\bar{v}\sigma^2/d+1}\right)^{1/2}+O\left(\frac{d^2}{\sigma^2}\right)\right)e^{-\frac{\pi}{4}\frac{\alpha_0^2}{(d/\sigma^2+\bar{v})^2}\left(\frac{d}{\sigma}\right)^2}+O\left(|t|\frac{d^2}{\sigma^2}+1\right)e^{-\frac{\pi}{4}\frac{d^2}{\sigma^2}}+O\left(e^{-\frac{\pi}{2}\sigma^2}\right)\nonumber\\
&=O\left(\frac{|t|}{\tau^{{\rm EC}_d}}e^{-\frac{\pi}{4} d^{\frac{\eta}{8}}}\right)
\end{align}
as defined in \cite[Eq.\ (F38)]{woods2018quantum} where $\sigma=d^{\eta/2}$ (cf.\ \cite[Eq.\ (F241)]{woods2018quantum}), $\bar{v}\sigma=d^{1-\eta/16}$ (cf.\ \cite[Eq.\ (F213)]{woods2018quantum}), and $\alpha_0$ can be set to one (cf.\ \cite[Eq.\ (F30)]{woods2018quantum}),
\begin{align}
\epsilon_{\rm tail}=O\left(e^{-\frac{\pi}{2}d^\eta}\right)
\end{align}
as defined in \cite[Eqs.\ (F78) and (F81)]{woods2018quantum},
\begin{align}
\Delta_{\rm EC}(t):=\sum_{k\in I_\gamma(s)}e^{-2\delta\int_k^{k+td/\tau^{{\rm EC}_d}}\d y\bar{V}_d(y)}g_{\rm nor}\left(k-\frac{sd}{\tau^{{\rm EC}_d}}\right)
\end{align}
where $I_\gamma(s):=\left\{\lfloor sd/\tau^{{\rm EC}_d}-\gamma d/2\rfloor,\dots,\lceil sd/\tau^{{\rm EC}_d}+1+\gamma d/2\rceil\right\}$. Here $g_{\rm nor}$ is a normal distribution satisfying $\sum_{k\in I_\gamma(s)}g_{\rm nor}\left(k-\frac{sd}{\tau^{{\rm EC}_d}}\right) \ge 1-\epsilon_{\rm tail}$.
\end{lemma}

The next step is to bound the dominant term $\Delta_{\rm EC}(t)$.
By \cite[Eq. (F13)]{woods2018quantum}, $x_{\rm vr}$ is defined so that
\begin{align}
1-\tilde{\epsilon}_V=\int_{-x_{\rm vr}}^{x_{\rm vr}}\d x\,\bar{V}_0(x+x_0),
\end{align}
where $\bar{V}_0$ is defined via the relation $\bar{V}_d(x)=\frac{2\pi}{d}\bar{V}_0\left(\frac{2\pi}{d}x\right)$ and $\tilde{\epsilon}_V$ is the vanishing term defined in Lemma \ref{lemma-quasiideal}. The relation between $\bar{V}_0$ and $\bar{V}_d$ implies that
\begin{align}
\int_k^{k+\frac{td}{\tau^{{\rm EC}_d}}}\d y\,\bar{V}_d(y)=\int_{\frac{2\pi k}{d}-x_0}^{\frac{2\pi k}{d}-x_0+\frac{2\pi t}{\tau^{{\rm EC}_d}}}\d x\,\bar{V}_0(x+x_0). 
\end{align}
Here we take the location of the potential to be $x_0=\pi$.
Then we can see that:
\begin{enumerate}
\item A sufficient condition for $\int_k^{k+td/\tau^{{\rm EC}_d}}\d y\bar{V}_d(y)\le \tilde{\epsilon}_V$ to hold is that
\begin{align}
\left[-x_{\rm vr},x_{\rm vr}\right]\subset\left[\frac{2\pi k}{d}-\pi,\frac{2\pi k}{d}-\pi+\frac{2\pi t}{\tau^{{\rm EC}_d}}\right]^c
\end{align}
holds for every $k\in I_\gamma(s)$, which is guaranteed when
\begin{align}
t\le \frac{\tau^{{\rm EC}_d}}2-s-\left(\frac{x_{\rm vr}}{2\pi}+\frac{\gamma}{2}\right)\tau^{{\rm EC}_d}=I_-.
\end{align}
\item A sufficient condition for $\int_k^{k+td/\tau^{{\rm EC}_d}}\d y\bar{V}_d(y)\ge 1-\tilde{\epsilon}_V$ to hold is that
\begin{align}
\left[-x_{\rm vr},x_{\rm vr}\right]\subset\left[\frac{2\pi k}{d}-\pi,\frac{2\pi k}{d}-\pi+\frac{2\pi t}{\tau^{{\rm EC}_d}}\right]
\end{align}
holds for every $k\in I_\gamma(s)$, which is guaranteed when
\begin{align}
t\ge \frac{\tau^{{\rm EC}_d}}2-s+\left(\frac{x_{\rm vr}}{2\pi}+\frac{\gamma}{2}\right)\tau^{{\rm EC}_d}=I_+.
\end{align}
\end{enumerate}

The above discussion yields the bounds for $\Delta_{\rm EC}(I_{-})$ and $\Delta_{\rm EC}(I_{+})$. Explicitly, we have:
\begin{align}
\Delta_{\rm EC}(I_{-})&\ge \left(\min_{k\in I_\gamma(s)}e^{-2\delta\int_{k}^{k+td/\tau^{{\rm EC}_d}}\d y\bar{V}_d(y)}\right)\cdot (1-\epsilon_{\rm tail})\\
&\ge e^{-2\delta\tilde{\epsilon}_V}\cdot(1-\epsilon_{\rm tail}),
\end{align}
and
\begin{align}
\Delta_{\rm EC}(I_+)&\le \left(\max_{k\in I_\gamma(s)}e^{-2\delta\int_k^{k+td/\tau^{{\rm EC}_d}}\d y\bar{V}_d(y)}\right)\\
&\le e^{-2\delta(1-\tilde{\epsilon}_V)}.
\end{align}
Therefore, we have
\begin{align}
\Tr\left[\rho(I_{-})\right]-\Tr\left[\rho(I_{+})\right]&\ge e^{-2\delta\tilde{\epsilon}_V}\cdot(1-\epsilon_{\rm tail})-e^{-2\delta(1-\tilde{\epsilon}_V)}-2\epsilon_{\rm tail}-\epsilon_{\rm trans}\left(I_-\right)-\epsilon_{\rm trans}\left(I_+\right)\\
&\ge 1-2\delta\tilde{\epsilon}_V-e^{-\delta}-3\epsilon_{\rm tail}-2\epsilon_{\rm trans}\left(\tau^{{\rm EC}_d}\right),
\end{align}
 having assumed that $\tilde{\epsilon}_V\le 1/2$ (which always holds since we consider only the asymptotics). One can easily see from the above equation that the tail probability $\epsilon^{{\rm EC}_d}$ is bounded as Eq. (\ref{error-clock}).

\end{widetext}

\end{document}